\documentclass[twocolumn, showpacs, showkeys, amsmath, amssymb, superscriptaddress]{revtex4-1}

\newif\iffinal
\finaltrue
\usepackage[normalem]{ulem} 	%need for strikethrough
\usepackage[usenames,dvipsnames]{color}
\usepackage{multirow}
\usepackage{hhline}
\usepackage{titlesec} % to change the section title, documentation: http://tug.ctan.org/macros/latex/contrib/titlesec/titlesec.pdf

\usepackage[cleanup={}]{pstool}
\usepackage{ifpdf}
\ifpdf\else
  \makeatletter\providecommand\href@noop{\@secondoftwo}\makeatother
\fi
\usepackage{dsfont}  			%%% for blackboard numbers and characters 
\usepackage{sansmath} %package to write symbols sanf serif in math environment

\usepackage{hyperref}
\usepackage[figure]{hypcap}
\newcommand{\del}[1]{\sloppy{\textcolor{blue}{\sout{#1}}}} % delete this
\newcommand{\macom}[1]{{\marginpar{\textcolor{blue}{#1}}}} % comment on margin without text highlight

\iffinal
	\renewcommand{\del}[1]{}

	\renewcommand{\macom}[1]{}
	
\else
\fi 

\newcommand{\executeiffilenewer}[2]{\ifnum\pdfstrcmp{\pdffilemoddate{./graphics/#2/#2.pdf}}{\pdffilemoddate{./graphics/#2/#2.eps}}>0{\includegraphics[width=#1]{./graphics/#2/#2} pdf newer! }\else{\psfragfig*[width=#1]{./graphics/#2/#2} pdf older!}\fi}

\renewcommand{\thesection}{\Roman{section}}
\titleformat{\section}{\center\bfseries\uppercase}{\thesection:}{1em}{} %needs package titlesec

\begin{document}

%////////////////
\title{Electronic Decoherence of Two-Level Systems in a Josephson Junction}
%\title{Two-Level-Systems in metal-insulator interfaces: Electronic Decoherence of Two-Level-Systems in a Josephson junction} %not good yet
%\title{Quasiparticle-Induced Decoherence of Microscopic Two-Level-Systems in a Superconducting Quantum Circuit}
%\title{Evidence of Interaction between Quasiparticles and Two-Level-Systems in a Superconducting Josephson Junction based Qubit}

\author{Alexander Bilmes}
	\affiliation{Physikalisches Institut, Karlsruhe Institute of Technology, 76131 Karlsruhe, Germany}
\author{Sebastian Zanker}
	\affiliation{Institut f\"ur Theoretische Festk\"orperphysik, Karlsruhe Institute of Technology, 76131 Karlsruhe, Germany}
\author{Andreas Heimes}
	\affiliation{Institut f\"ur Theoretische Festk\"orperphysik, Karlsruhe Institute of Technology, 76131 Karlsruhe, Germany}
\author{Michael Marthaler}
	\affiliation{Institut f\"ur Theoretische Festk\"orperphysik, Karlsruhe Institute of Technology, 76131 Karlsruhe, Germany}	
\author{Gerd Sch\"on}
	\affiliation{Institut f\"ur Theoretische Festk\"orperphysik, Karlsruhe Institute of Technology, 76131 Karlsruhe, Germany}
\author{Georg Weiss}
	\affiliation{Physikalisches Institut, Karlsruhe Institute of Technology, 76131 Karlsruhe, Germany}
\author{Alexey V. Ustinov}
	\affiliation{Physikalisches Institut, Karlsruhe Institute of Technology, 76131 Karlsruhe, Germany}
	\affiliation{Russian Quantum Center, National University of Science and Technology MISIS, Moscow 119049, Russia}
\author{J\"urgen Lisenfeld}
	\affiliation{Physikalisches Institut, Karlsruhe Institute of Technology, 76131 Karlsruhe, Germany}
	
\date{\today}

\begin{abstract}
The sensitivity of superconducting qubits allows for spectroscopy and coherence measurements on individual two-level systems present in the disordered tunnel barrier of an $\mathrm{Al/AlO_x/Al}$ Josephson junction. We report experimental evidence for the decoherence of two-level systems by Bogoliubov quasiparticles leaking into the insulating $\mathrm{AlO_x}$ barrier. We control the density of quasiparticles in the junction electrodes either by the sample temperature or by injecting them using an on-chip dc-SQUID driven to its resistive state. The decoherence rates were measured by observing the two-level system's quantum state evolving under application of resonant microwave pulses and were found to increase linearly with quasiparticle density, in agreement with theory. This interaction with electronic states provides a noise and decoherence mechanism that is relevant for various microfabricated devices such as qubits, single-electron transistors, and field-effect transistors. The presented experiments also offer a possibility to determine the location of the probed two-level systems across the tunnel barrier, providing clues about the fabrication step in which they emerge.
%Equally, it provides information about TLS' position across the Josephson junction: TLS may emerge in the grown oxide on a Josephson junction's bottom electrode rather than in the interface of this oxide and the top electrode.
\end{abstract}

\maketitle 

\setlength{\parskip}{-0.25cm}

%--------------------SECTION----------------------------------------
\section{Introduction}
\label{sec_Intro}
While superconducting circuits based on Josephson junctions (JJs) rapidly mature towards favorable and applicable qubits for quantum computers~\cite{ClarkeWilhelmReview,Barends2013, Kelly2015}, a major source of their decoherence traces back to spurious material defects that give rise to the formation of low-energy two-level systems (TLSs). On the other hand, sensitivity to tiny perturbations turns JJ qubits into ideal tools to study the properties of TLSs. For example, microwave spectroscopy of JJ phase qubits shows avoided level crossings revealing the TLSs' quantum character as well as their coherent interaction with the qubit \cite{TLSOscis}. Various microscopic models including dangling bonds, Andreev bound states~\cite{Faoro2005}, and Kondo fluctuators~\cite{Ansari2011} have been suggested to explain the origin of TLSs. There is growing evidence~\cite{DuBois:NJP:2015,Rubio:PRB:2014}, however, that they are formed by small groups of atoms that are able to tunnel between two energetically almost equivalent configurations. This is most strongly supported by recent experiments where the TLSs' energy splittings were tuned by applying external static strain~\cite{Grabovskij:S:2012}. TLSs are the source of low-energy excitations, which are also responsible for the thermal, acoustic, and dielectric properties of glasses at temperatures below $1\,\mathrm{K}$~\cite{Phillips:JLTP:1972,Anderson:PM:1972}, which are well studied in bulk materials. Inherent to disordered solids, they are present in surface oxides and insulating layers of any microfabricated device as well as in the tunnel barriers of Josephson junctions.\\

In contrast to traditional measurements performed on glasses that probe huge ensembles of TLSs, the sensitivity of JJ-based qubits allows one to address \textit{single} TLSs and determine their individual properties. Strain-tuning experiments, e.g., measure a TLS's deformation potential~\cite{Grabovskij:S:2012} and allow for a detailed analysis of the coherent interaction between two TLSs brought into resonance~\cite{Lisenfeld2015}. In another experiment, the temperature dependence of energy-relaxation and dephasing rates of individual TLSs were measured~\cite{Lisenfeld:PRL:2010} - with an unexpected and yet unexplained result: The energy-relaxation rate $\Gamma_1$ increased much more rapidly with temperature than predicted by the one-phonon scattering process dominating in dielectric solids~\cite{Jaeckle72}.\\

Earlier work showed that in metallic hosts, inelastic scattering of conduction electrons ~\cite{Black81} may outweigh the phonon-induced $\Gamma_1$ at sufficiently low temperatures. This process was verified in ultrasonic absorption and phonon echo experiments for TLS in superconducting metallic glasses~\cite{Weiss80,Weiss88} as well as for hydrogen TLSs in niobium~\cite{Morr89}. In the superconducting state, an energy gap opens and the electronic excitations are Bogoliubov quasiparticles (QPs). In ideal BCS systems, their density decreases below the superconducting transition $T_\text{c}$ and accordingly the electron-induced TLS relaxation falls off by several orders of magnitude. On the other hand, thermally excited QPs as well as so-called excess QPs, which may stem from stray infrared photons~\cite{Echternach2008} or other unknown sources, may still lead to TLS relaxation below $T_\text{c}$.\\

In this paper, we report on experimental studies of the dynamics of TLSs residing in the amorphous insulating barrier of a JJ (i.e., junction TLSs) and present evidence for their interaction with QPs whose evanescent wave functions leak from the superconducting Al film into the insulator. The density of QPs is controlled by two complementary methods: either by injecting QPs with an on-chip dc-SQUID~\cite{Lenander2011} at a constant mixing-chamber temperature of $30\,\mathrm{mK}$ or by variation of temperature up to $330\,\mathrm{mK}$. In this temperature range, the contribution of phonons to the decoherence of the TLS with energy splitting comparable to $k_\text{B}\text{T}$ can be regarded as almost constant~\cite{Jaeckle72,Lisenfeld:PRL:2010}. To observe the TLS' quantum state evolution, we drive them directly using protocols of resonant microwave pulses, while the qubit is only operated for TLS readout~\cite{Lisenfeld:PRL:2010}. Further, a piezoactuator transfers mechanical strain to the sample and controls the TLS asymmetry energy $\varepsilon$ via its elastic dipole moment (see Appendix \ref{sec_piezo} for technical details). This strain tuning~\cite{Grabovskij:S:2012} enables us to explore the TLS response to QPs for varying $\varepsilon$.\\

\begin{figure}[tbph]
	\begin{minipage}{0.49\textwidth}
		\includegraphics[width=\linewidth]{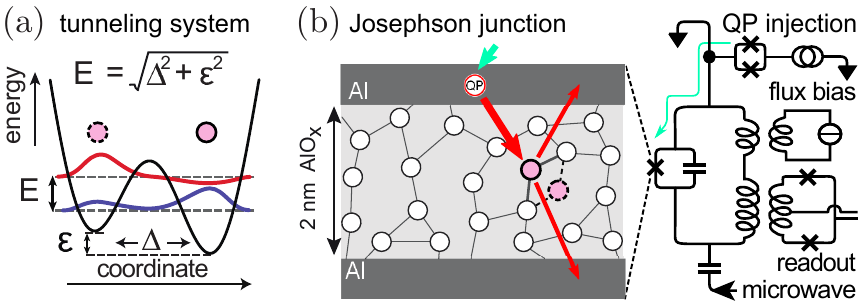}{
%		\psfragfig[width=\linewidth]{./graphics/jjQubitDoublewell/jjQubitDoublewellTunnel}{
%		\psfrag{(a)}{(a)}
%		\psfrag{(b)}{(b)}
%		\psfrag{(c)}{(c)}
		}
	\end{minipage}
	\label{fig_Doublewell}
	\caption{(a) The double-well potential of a TLS, where $E$ is the TLS transition energy composed by the TLS' tunneling energy $\mathit{\Delta}$ and its asymmetry energy $\varepsilon$. (b) Right: Schematic of the phase qubit circuit. The green arrow indicates the diffusion of quasiparticles from the injector SQUID to the qubit junction. Left: Sketch of a JJ depicting a TLS. The red arrows show scattering of QPs on a TLS: backscattering into the initial electrode or scattering into the opposite electrode.}
\end{figure}

%--------------------SECTION----------------------------------------
\section{Model}
\label{sec_Model}

Within the standard tunneling model~\cite{Phillips:JLTP:1972,Anderson:PM:1972}, TLS are described as virtual particles bound in a double-well potential as illustrated in Fig.~\ref{fig_Doublewell}(a), where the left and right wells correspond to one or another metastable TLS atomic configuration. The TLS' energy scale is given by the tunneling energy $\mathit{\Delta}$ and the asymmetry energy $\varepsilon$. The unperturbed TLS Hamiltonian reads
\begin{equation}
H=\frac{1}{2}\mathit{\Delta}\sigma_x + \frac{1}{2}\varepsilon\sigma_z \equiv \frac{1}{2}E\tau_z,
\label{eq_TLSHam}
\end{equation}
where $\sigma_x$ and $\sigma_z$ are Pauli matrices. The transition energy is $E=\sqrt{\mathit{\Delta}^2+\varepsilon^2}$ and $\tau_z$ is the Pauli matrix in the diagonalized or energy basis.\\

TLSs couple to elastic and electric fields by respective dipole moments, predominantly varying the asymmetry energy $\varepsilon$. In the energy basis, this coupling gives rise to longitudinal ($\propto \tau_z$) and transversal ($\propto \tau_x$) perturbation terms. The latter describes transitions between the energy eigenstates and explains, e.g., the one-phonon relaxation and, in particular, the resonant coupling of the junction TLS to the JJ qubit via the electric field within the junction, which enables readout and coherent manipulation of TLS quantum states~\cite{TLSOscis}. The TLS interaction with electrons of a metallic environment arises from inelastic scattering of the electrons and is expressed as
\begin{equation}
H_\text{el}=\sigma_{z}\sum_{k,k',\sigma}g_{k,k'}c^\dagger_{k,\sigma}c^{\phantom{(m)}}_{k',\sigma},\label{eq_HamElTLS}
\end{equation}
where the summation runs over the spin degree of freedom $\sigma$ and the electronic eigenstates $k$, $k'$ that are not necessarily plane waves. The scattering matrix elements are designated by $g_{k,k'}$. The presence of $\sigma_z$ in $H_\text{el}$ indicates that electrons experience a change in the scattering potential depending on the two configurations of the TLS~\cite{Black81}. Rewriting $\sigma_z$ in the energy basis and introducing the averaged scattering matrix $g$, we obtain
\begin{align}
H_\text{el} = g\left(\frac{\mathit{\Delta}}{E}\tau_x + \frac{\varepsilon}{E}\tau_z\right) \sum_{k,k',\sigma}c^\dagger_{k,\sigma}c^{\phantom{(m)}}_{k',\sigma}.\label{eq_HamElTLS2}
\end{align}
In Ref.~\cite{Black81}, the averaged transversal and longitudinal scattering matrix elements, $V_\perp\equiv gN\mathit{\Delta}/E$ and $V_\parallel\equiv gN\varepsilon/E$, are reported to have a magnitude up to $0.1\,\mathrm{eV}$, where $N$ is the number of atoms in the system. The probed TLSs reside in the insulating barrier of a JJ; thus we estimate $N$ $\approx 10^9$ from the volume of the tunnel barrier ($1\,\mathrm{\mu m^2}\times 2\,\mathrm{nm}$) and a typical atomic volume ($10^{-30}\,\mathrm{m}^3$). Thus, $g$ scales as $V_\perp/N=10^{-10}\,\mathrm{eV}$. In the superconducting state, the electronic excitations are obtained after a Bogoliubov transformation. Hence, $H_\text{el}$ turns into
\begin{align}
H_\text{QP}=\;&g\left(\frac{\mathit{\Delta}}{E}\tau_x + \frac{\varepsilon}{E}\tau_z\right)\nonumber\\
&\times\sum_{k,k',\sigma,l}s_{l}(u^{(l)}_{k}u^{(l)}_{k'}-v^{(l)}_{k'}v^{(l)}_{k})\alpha^{(l)\dagger}_{k,\sigma}\alpha^{(l)}_{k',\sigma},
\label{eq_HamQPTLSJJ}
\end{align}\\
where $u$ and $v$ are BCS real-number coherence factors. Further, $l=B,T$ indicates the bottom and top electrodes of the JJ. The probed TLSs reside in the insulating barrier of a JJ; accordingly, only the leaky portions of the QPs' wave functions from the electrodes are to be taken into account. We thus have introduced in Eq.~\eqref{eq_HamQPTLSJJ} the averaged probability $s_l$ for a QP to interact with a TLS and to return into the initial electrode. It decays exponentially with the distance between the electrode and the TLS. Processes where a QP is scattered to the opposite electrode only weakly contribute to the QP-TLS interaction and are neglected in $H_\text{QP}$ (see the full Hamiltonian in Appendix \ref{sec_TLSPosition}).\\

The QP-induced energy-relaxation rate of the TLS is calculated from Eq.~\eqref{eq_HamQPTLSJJ} using Fermi's golden rule~\cite{Zanker2016}:
\begin{align}
%\Gamma_1=&M_1^2(s_\text{B}^2\Gamma_{1}^{\text{(L)}} + s_{R}^2\Gamma_{1}^{\text{(T)}}),\label{eq_gamma1TLS}\\
\Gamma_1=&\;s_\text{B}^2\Gamma_{1}^{\text{(B)}} + s_\text{T}^2\Gamma_{1}^{\text{(T)}},\label{eq_gamma1TLS}\\
\Gamma_1^{(l)}=&\;\frac{4\pi}{\hbar}\left(N_0Vg\frac{\mathit{\Delta}}{E}\right)^2\mathit{\Delta}_\text{s}\int_{1}^{\infty}d\epsilon\left(1-\frac{1}{\epsilon(\epsilon+E/\mathit{\Delta}_\text{s})}\right)\notag\\
&\times \rho(\epsilon)\rho(\epsilon+\frac{E}{\mathit{\Delta}_\text{s}}) f_0^{(l)}(\epsilon)\left(1-f_0^{(l)}(\epsilon+\frac{E}{\mathit{\Delta}_\text{s}})\right).\label{eq_gamma1TLS_B}
\end{align}
The electronic density of states at the Fermi edge including the spin degeneracy is given by $2N_0$ and the reduced QP density of states is $\rho(\epsilon)=\epsilon/\sqrt{\epsilon^2-1}$, where $\epsilon$ is the QP energy in units of the BCS gap $\mathit{\Delta}_\text{s}$ in Al. The integral in Eq.~\eqref{eq_gamma1TLS_B} takes into account all possible absorption processes where a QP scatters from a state of energy $\epsilon$ into that of energy $\epsilon+E$. We approximate the QP distribution function on each electrode by the Fermi function $f_0^{(l)}$ [see explanations of Eq.~\eqref{eq_gamma1} in \ Appendix \ref{sec_QPInject}]. The probe volume $V$ is estimated to be of the order of the cubic electron coherence length in aluminum of $1\,\mathrm{\mu m}^3$. Another relevant rate is $\Gamma_\text{R}$, the decay rate of Rabi oscillations in situations when the TLS is continuously driven. $\Gamma_\text{R}$ follows from Eq.~\eqref{eq_gamma1TLS} after the substitution $E\rightarrow \Omega$ in the integrand of Eq.~\eqref{eq_gamma1TLS_B}, where $\Omega\approx h\cdot10\,\mathrm{MHz}$ is the typical coupling strength of the driving microwave to the probed TLS. The pure dephasing rate $\Gamma_2^*$ is derived from Eq.~\eqref{eq_gamma1TLS} by replacing in the prefactor $\mathit{\Delta}/E\rightarrow \varepsilon/E$ and by setting $E\rightarrow 0$ in the integrand.

%--------------------SECTION----------------------------------------
\section{Experimental Results}
\label{sec_Results}
\begin{figure}[b!]
	\begin{minipage}{0.49\textwidth}
		\vspace{0pt}
%		\psfragfig*[width=\linewidth]{./graphics/qubxqp/qubxqp} % does not work with citations with additional text included
		\includegraphics[width=\linewidth]{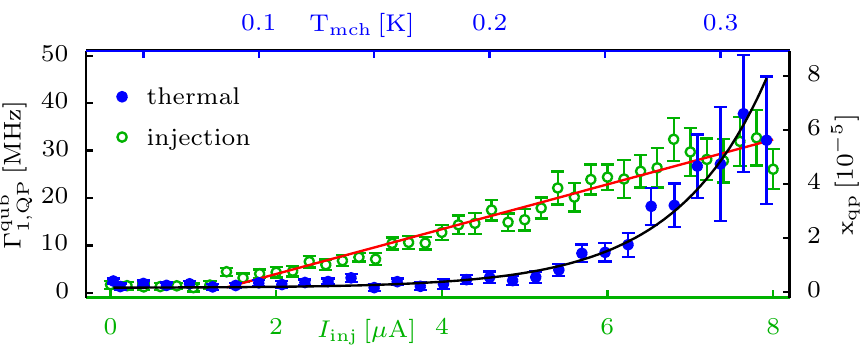}
	\end{minipage}
	\label{fig_G1_vs_QP}
	\caption{Quasiparticle-induced energy-relaxation rate of the qubit $\gamma_{1}^\text{qub}$ (left axis) recorded in two complementary experiments: increasing the mixing-chamber temperature $T_\text{mch}$ (top axis) and applying current $I_\text{inj}$ to the injector dc-SQUID (bottom axis). The thermally generated quasiparticle density significantly increases for $T_\text{mch}>200\,\mathrm{mK}$, while the injection of quasiparticles starts when $I_\text{inj}$ exceeds the SQUID's switching current of $1.5\,\mathrm{\mu A}$. The quasiparticle density $x_\text{qp}$ (right axis) is numerically deduced from $\gamma_{1}^\text{qub}$ (see Appendix \ref{sec_QPInject} for details). Both the linear fit (red line) and the exponential fit (black line) are used for calibration of $x_\text{qp}$ vs $I_\text{inj}$ and $T_\text{mch}$, respectively.}
\end{figure}

The normalized QP density is defined as
\begin{equation}
x_{\text{qp}}\equiv\frac{n_\text{qp}}{2\mathit{\Delta}_\text{s} N_0}=\int_1^\infty d\epsilon\rho(\epsilon)f_0(\epsilon,T,\mu),
\label{eq_qp}
\end{equation}
where $n_\text{qp}$ is the total QP density and $2\mathit{\Delta}_\text{s} N_0$ is the Cooper pair density at zero temperature. The QP partition function $f_0(\epsilon,T,\mu)$ depends on the QP temperature $T$ and the chemical potential $\mu$. As mentioned before, we control the QP density either via the mixing-chamber temperature $T_\text{mch}$ or by QP injection that shifts $\mu$. In the latter method, we use an injector dc-SQUID that is galvanically coupled to the JJ via a common thin-film Al ground plane [see Fig.~\ref{fig_Doublewell}(b)]. Following Ref.~\cite{Lenander2011}, we apply bias current pulses (of amplitude $I_\text{inj}$) to the injector dc-SQUID exceeding its switching current to produce QPs from Cooper pair breaking processes, which then diffuse over a distance of $1\,\mathrm{mm}$ through the ground plane towards the JJ. We performed measurements of $x_\text{qp}$ for varying delays after the start of QP injection and found good agreement with results from simulations of QP diffusion in a simplified two-dimensional geometry (see Appendices \ref{sec_QPInject} and \ref{sec_QPDiff}). Further, we show in Appendix \ref{sec_QPHeating} the analysis of switching current statistics of the readout-SQUID, with which we verify that the QP injection does not heat the sample. In both the thermal and the injection experiments, we controlled $x_{\text{qp}}$ by monitoring the QP-induced energy-relaxation rate $\gamma_{1}^\text{qub}$ of the qubit as a function of the mixing-chamber temperature $T_\text{mch}$ and $I_\text{inj}$, respectively (see Fig.~\ref{fig_G1_vs_QP}). From $\gamma_{1}^\text{qub}$ we deduced the value of $x_{\text{qp}}$ that is plotted on the right vertical axis~\cite{Catelani2011} (see Appendix \ref{sec_QPInject} for details). The continuous lines are the corresponding fits, which provide the calibration of $x_\text{qp}$ vs each $I_\text{inj}$ and $T_\text{mch}$ that are used for quantitative comparison of the TLS relaxation in the thermal and injection experiments.\\
\begin{figure} %[b!]
	\begin{minipage}{0.49\textwidth}
%		\psfragfig*[width=\linewidth]{./graphics/fittlsnqp/fittlsnqp} % does not work with citations with additional text included
		\includegraphics[width=\linewidth]{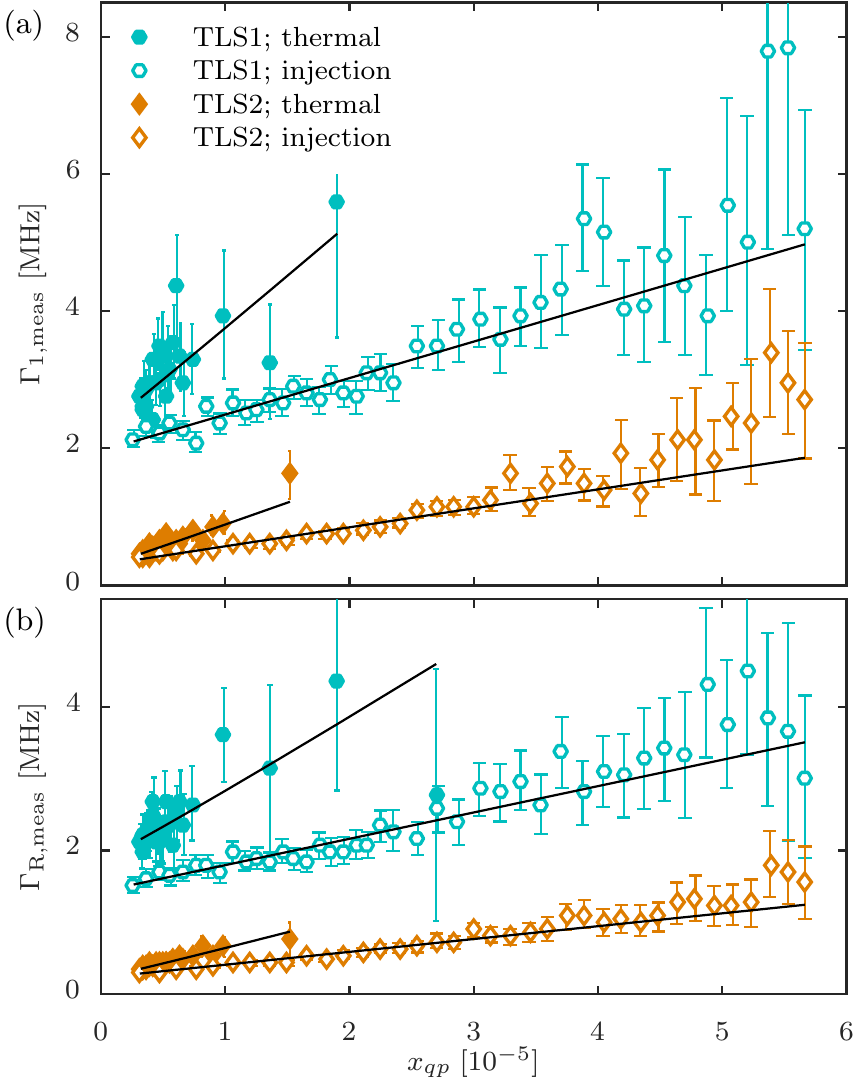}		
	\end{minipage}
	\label{fig_fittlsnqp}
	\caption{ (a) Measured energy-relaxation rates $\Gamma_{1,\text{meas}}$ of two distinct TLSs vs quasiparticle density $x_\text{qp}$~\cite{[{Measurements on injection of quasiparticles with similar results are reported in }][{ (unpublished)}]Bilmes_Dipl2014}. The legends indicate TLS labels and whether quasiparticles were injected or thermally generated. Below $330\,\mathrm{mK}$, the temperature dependence of the phonon-induced TLS decoherence is negligible. Thus, we fit our data to the purely QP-induced decoherence rate $K^m\cdot\Gamma_{1}^{\text{(B)}}+const$ shown in Eq.~\eqref{eq_gamma1TLS_B} (black lines). The corresponding fit factors $K^m$ are indexed with the type of QP generation and listed in Table~\eqref{tabK}. In Appendix \ref{sec_TLSPosition}, we present these data in a double-logarithmic plot, which is more readable at low $x_\text{qp}$. (b) Recorded decay rates $\Gamma_\text{R,meas}$ of TLS Rabi oscillations vs $x_\text{qp}$ and the corresponding fits.}
\end{figure}
\begin{table}
\centering
\begin{tabular}{ |l||l|l|l|l|l| }
\hhline{~-----}
\multicolumn{1}{l|}{\multirow{2}{*}{}} & \multicolumn{1}{c|}{$\mathit{\Delta}$} & \multicolumn{2}{|c|}{energy relax. $\Gamma_{1,\text{meas}}$} & \multicolumn{2}{|c|}{Rabi decay $\Gamma_\text{R,meas}$} \\\hhline{~|~|----}
\multicolumn{1}{l|}{} & \multicolumn{1}{l|}{$(h\mathrm{\cdot GHz})$} & $K^{\text{th}}$\hspace{9pt} & $K^{\text{inj}}$ & $K^{\text{th}}$ & $K^{\text{inj}}$ \\\hhline{-|=====|} %\hhline{-||-----}
TLS1 & 6.219 & 0.32 & 0.12 & 0.30 & 0.12 \\ \hhline{|~||~~~~~}
TLS2 & 6.667 & 0.14 & 0.06 & 0.13 & 0.06 \\ \hhline{|-|-----}
\end{tabular}
\caption{Tunnelling energies $\mathit{\Delta}$ of the probed TLSs and factors $K^m$ from the fits [see Fig.~\eqref{fig_fittlsnqp}] to the measured TLS' energy-relaxation rate $\Gamma_{1,\text{meas}}$ and the decay rate of Rabi oscillations $\Gamma_\text{R,meas}$.}
\label{tabK}
\end{table}

TLSs are excited by resonant microwave pulses applied to the circuit, while the qubit is detuned by about $1\,\mathrm{GHz}$ from the transition frequency of the probed TLS. For TLS readout, we tune the qubit by a short rectangular flux pulse into resonance with the TLS in order to swap their quantum states, followed by qubit readout. The TLS decoherence rates were obtained using standard measurement protocols that have been established in earlier work~\cite{Lisenfeld:PRL:2010}. In Fig.~\ref{fig_fittlsnqp}, we present the response of two distinct TLSs to QPs, whose tunneling energies $\mathit{\Delta}$ are listed in Table~\eqref{tabK}, while their asymmetries were strain tuned close to zero~\cite{Grabovskij:S:2012} (see  Appendix \ref{sec_TLSPosition} for data at further $\varepsilon$ values). The measured energy-relaxation rate $\Gamma_\mathrm{1,meas}$ and Rabi decay rate $\Gamma_\mathrm{R,meas}$ are plotted as a function of $x_\text{qp}$. The black lines are fits of $K^m\cdot\Gamma_\text{1}^{\text{(B)}}+const$ and $K^m\cdot\Gamma_\text{R}^{\text{(B)}}+const$ to the experimental data in Figs.~\ref{fig_fittlsnqp}(a) and ~\ref{fig_fittlsnqp}(b), respectively. Here, $K^m\equiv s_\text{B}^2$ is treated as a fit parameter as listed in Table~\eqref{tabK}, where $m=\text{th},\text{inj}$ designates whether QPs were thermally generated or injected. The constant contribution originates from the coupling to phonons and neighboring TLSs. Apparently, fits to the theoretical prediction from Eq.~\eqref{eq_gamma1TLS_B} describe our data very well. In particular, for a given TLS, we extract the same values $K^m$ from fits to $\Gamma_1$ and $\Gamma_\text{R}$, respectively. By simplifying the integral in Eq.~\eqref{eq_gamma1TLS_B}, one finds a linear dependence of TLS decoherence rates on QP density confirmed by the fit: $\Gamma_1,\Gamma_\text{R}\propto x_\text{qp}$. The fit parameters have an average magnitude of about $0.1$ that traces from the exponential decay of the QP wave-function within the tunnel barrier. Accordingly, in a JJ, we state the scattering matrix element $V_\perp$ to be of the order of $10\,\mathrm{meV}$.\\

The pre factor $\mathit{\Delta}^2/E^2=\mathit{\Delta}^2/(\mathit{\Delta}^2+\varepsilon^2)$ in Eq.~\eqref{eq_gamma1TLS_B} includes the dependence of QP-induced energy-relaxation and Rabi decay rates on the TLS' asymmetry energy $\varepsilon$. To verify this, we repeated the measurements after TLS1 was strain tuned to a large asymmetry energy $\varepsilon=3.299\,h\cdot\mathrm{GHz}$, corresponding to a reduction of $\mathit{\Delta}^2/E^2$ by $12\%$. However, since the confidence interval in determining $K^m$ was about $\pm 14\%$, we could not detect any significant strain dependence.
On the other hand, the QP-induced pure dephasing rate $\Gamma_2^*$ depends on the asymmetry energy as $\varepsilon^2/E^2$. In fact, we found that it vanishes at the TLS symmetry point ($\varepsilon\approx0$) and increases otherwise slightly with $x_\text{qp}$ (in Appendix \ref{sec_TLSDephasing}, we show the relevant data). However, for $\varepsilon\neq0$, the pure dephasing of the probed TLS is dominated by its interactions with thermally fluctuating TLS~\cite{Lisenfeld2016}.\\

We see in Fig.~\eqref{fig_fittlsnqp} that for fixed $x_\text{qp}$, thermally generated QPs always lead to stronger TLS' energy-relaxation than injected QPs. The $\varepsilon$-averaged ratio $K^\text{th}/K^\text{inj}$ for TLS1 and TLS2 is $2.5$ and $1.9$, respectively. This can be explained from the fact that $x_\text{qp}$ will increase equally in both JJ electrodes with increasing temperature, whereas injected QPs predominantly appear in the top electrode that is connected directly to the ground plane. We numerically solved the stationary Boltzmann equation and found the QP imbalance $\alpha\equiv x_\text{qp}^\text{(T)}/x_\text{qp}^\text{(B)}$ between top and bottom electrodes to be in the range of 2 to 4. Due to the fast exponential decrease of $s_{l}^2$ within the tunnel barrier, one of the two terms in Eq.~\eqref{eq_gamma1TLS} is dominant when the probed TLS is closer to one or the other electrode. Thus, a TLS residing near the bottom electrode would experience the presence of more QPs in the thermal experiment than in the injection experiment. Numerical and analytical calculations of the ratio $K^\text{th}/K^\text{inj}$ as a function of the TLS' location between the electrodes suggest that the probed TLSs are located closer to the bottom than to the top electrode (Appendix \ref{sec_TLSPosition}). Thus, it seems that in the $\mathrm{Al/AlO_x/Al}$ junctions used in this work~\cite{Steffen2006}, TLSs preferably emerge during the thermal oxidation of the Ar-milled bottom electrode rather than during the successive deposition of the top electrode. This assumption could be verified by repeating such experiments on a sample containing two identical qubits, whose JJs are connected to the ground plane either by their top or bottom electrodes, respectively. Alternatively, one could selectively inject QPs from both sides of the JJ.

%--------------------SECTION----------------------------------------
\section{Summary}
\label{sec_Summary}
In conclusion, we have explained the rapid increase of the energy-relaxation rates of two-level systems (TLSs) with temperature observed in previous work~\cite{Lisenfeld:PRL:2010}: TLSs that reside in the Josephson junctions' tunnel barrier of a qubit couple to the evanescent wave function of quasiparticles (QPs) in the electrodes. The TLS' energy-relaxation rate is proportional to the QP density and hence increases exponentially with temperature. In our experiments, the QP density was controlled either by varying the temperature of the sample or by injecting QPs using an on-chip dc-SQUID~\cite{Lenander2011}. The superconducting phase qubit served both as a monitor for the QP density and for TLS readout. Simulations of injected QPs diffusing towards the Josephson junction match the measured QP density during and after the QP injection pulse. We found good agreement between the theoretical prediction and the measured increase of the TLS' energy-relaxation and Rabi decay rates as a function of the QP density. Moreover, we found a difference in the strength of TLS decoherence comparing thermally generated to injected QPs, which we explain by the particular location of the TLS in the junction. Such measurements thus provide a possibility to determine in which fabrication step TLSs emerge.\\

These findings concern a variety of microfabricated circuits in which TLSs reside within native oxides or grown dielectric layers close to a conductor. The electron-TLS interaction analyzed here provides a mechanism of decoherence and fluctuations that may be relevant, e.g., for semiconductor devices such as gated quantum dots and field-effect transistors. Likewise, it can explain a reduction in mutual TLS coupling due to enhanced TLS relaxation rates as it was found in recent experiments where a superconducting resonator was capped by a normal conducting platinum layer~\cite{Burin2015,Burnett2016}.

%--------------------SECTION----------------------------------------
\section{Acknowledgements}
\label{sec_Acknowledgements}
We would like to thank J.M. Martinis (University of California Santa Barbara) for the qubit sample we have measured in this work. A.B. acknowledges financial support by the Graduate Funding (Landesgraduiertenf\"orderung) from the German States Program at Karlsruhe Institute of Technology that is funded by the Ministry of Science, Research and the Arts of the German State of Baden-W\"urttemberg, and by the Helmholtz International Research School for Teratronics (HIRST). Support by the Deutsche Forschungsgemeinschaft (DFG) (Grants LI2446/1-1 and SCHO287/7-1) is gratefully acknowledged. Partial support by the Ministry of Education and Science of Russian Federation in the framework of Increase Competitiveness Program of the NUST MISiS (Grant К2-2016-063) is gratefully acknowledged.\\

%---APPENDIX---------------------------------------------------
%--------------------------------------------------------------
%--------------------------------------------------------------
\renewcommand{\theequation}{A\arabic{equation}}
\setcounter{equation}{0}
\renewcommand{\thesection}{\Alph{section}}
\setcounter{section}{0}
\titleformat{\section}{\bfseries\uppercase}{APPENDIX \thesection:}{1em}{} %needs package titlesec

%--------------------SECTION----------------------------------------
\section{Strain-tuning of TLS}
\label{sec_piezo}
\begin{figure}[htbp]
	\begin{minipage}{0.24\textwidth}
%  	\psfragfig*[width=\linewidth]{./graphics/piezosketch/piezosketch}{
  	\includegraphics[width=\linewidth]{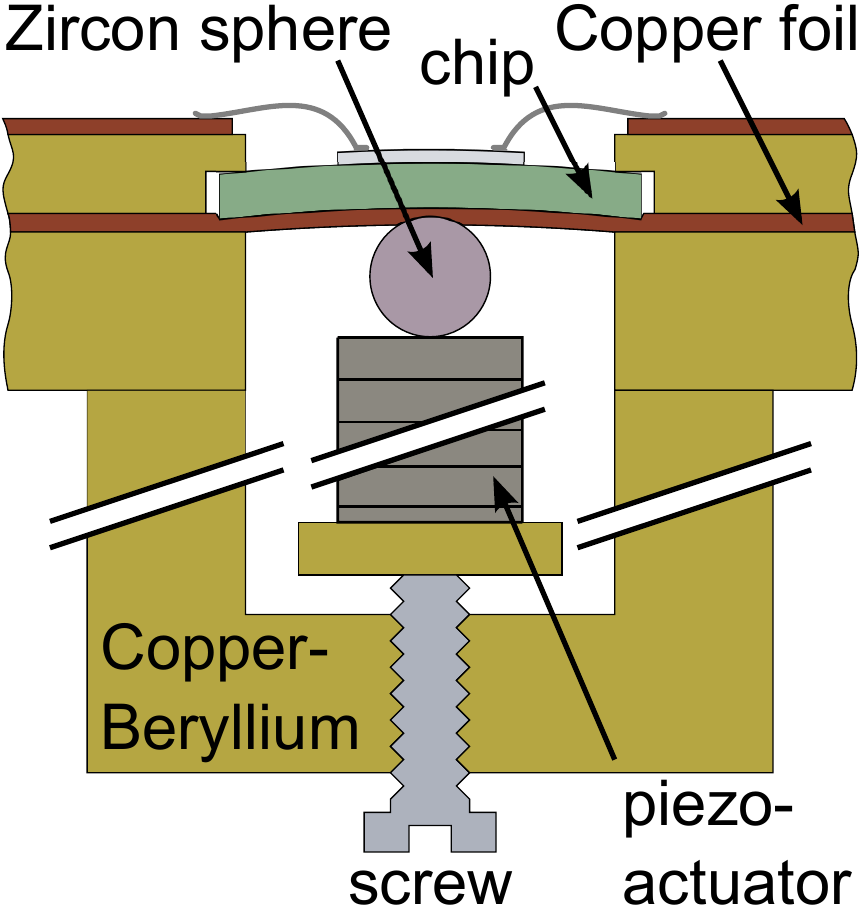}{
  	}
	\end{minipage}
	\begin{minipage}{0.24\textwidth}
%	\psfragfig*[width=\linewidth]{./graphics/piezomodel/piezomodel}{
%  	\includegraphics[width=\linewidth]{./graphics/piezomodel/piezomodel}{
  	\includegraphics[width=\linewidth]{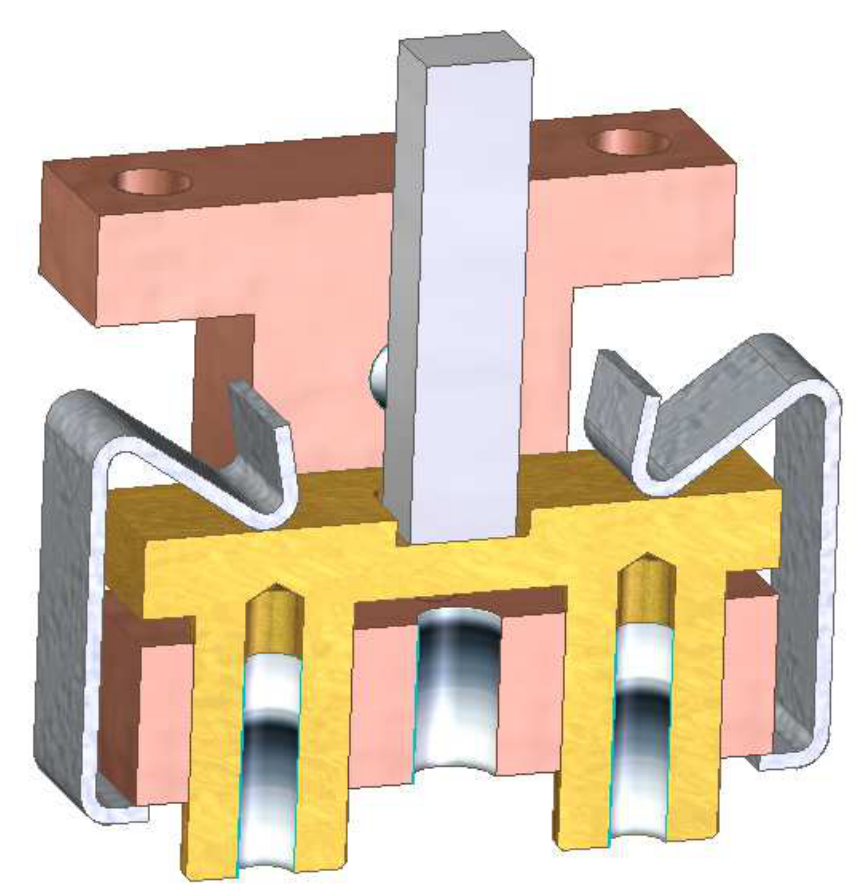}{
  	}
	\end{minipage}	
	\caption{Left: Sketch of the sample holder that enables us to change the TLS' asymmetry energy $\varepsilon$ using a piezoactuator that transfers elastic strain onto the chip with the qubit. The zircon sphere provides a one-point contact of the piezocrystal and the chip, while the Copper foil screens the electromagnetic crosstalk. Right: Cross section of the piezo-holder that consists of the main frame (brown, Cu-Be) and the slide (yellow, Cu-Be) that is held by two springs (Cu-Be). A brass screw that fits through the middle tapped whole adjusts the vertical position of the slide, while the piezocrystal (gray) is glued onto the slide.}
	\label{fig_piezo}
\end{figure}

At University of California Santa Barbara, the sample was microfabricated on a chip made of sapphire that is gripped in a sample holder (see Fig.~\ref{fig_piezo}), while at the bottom side a stack-piezoactuator~\cite{[{P-882.11 PICMA $\textregistered$ Piezoaktor $3\,\mathrm{mm}\times2\,\mathrm{mm}\times9\,\mathrm{mm}$, PI Ceramic GmbH, Lindenstraße, 07589 Lederhose, Germany}][{}]piezo} is mounted, whose elongation is controlled by the applied dc voltage $\mathrm{V_p}$. The transferred strain to the chip tunes the asymmetry energy of the TLS, $\varepsilon(\mathrm{V_p})=\gamma(\epsilon(\mathrm{V_p})-\epsilon_0)$, via its elastic dipole moment. Here, $\gamma=\partial \varepsilon/\partial\epsilon$ is the deformation potential that depends on the orientation of the TLS' elastic dipole moment relative to the elongation vector on the concave side of the chip. $\epsilon(\mathrm{V_p})\approx(\partial\epsilon/\partial\mathrm{V_p})\cdot\mathrm{V_p}$ is the effective strain field, while the coefficient $\partial\epsilon/\partial\mathrm{V_p}$ is estimated to $\approx10^{-7}/\mathrm{V}$ based on a measurement of the piezoelongation at a temperature of $4.2\,\mathrm{K}$ and finite element simulation of the resulting chip deformation~\cite{Grabovski_Thesis}.

%--------------------SECTION----------------------------------------
\section{Injection of Quasiparticles}
\label{sec_QPInject}
\begin{figure*}[htbp]
	\centering
	\begin{minipage}{0.8\textwidth}
		\centering
%		\psfragfig*[width=0.8\textwidth]{graphics/chip/chip}{
%			\psfrag{squid1}{\parbox{1cm}{\textcolor{white}{injector\\SQUID}}}
%			\psfrag{squid2}{\parbox{1cm}{\textcolor{white}{readout\\SQUID}}}
%			\psfrag{flux}{\textcolor{white}{ flux coil}}
%			\psfrag{ancilla}{\textcolor{white}{dummy coil}}
%			\psfrag{qub2}{\textcolor{white}{qubit coil}}
%			\psfrag{bottle}{\textcolor{white}{ "bottleneck"}}
%			\psfrag{short}{\textcolor{white}{ via}}
%			\psfrag{jos}{\textcolor{white}{ JJs}}
%			\psfrag{100}{\textcolor{white}{$\boldsymbol{100\mu m}$}}	
%			\psfrag{JJ}{ $\;\;$JJ}
%%	  	\includegraphics[width=0.8\textwidth]{graphics/chip/chip_arxiv}{
	  	\includegraphics[width=0.8\textwidth]{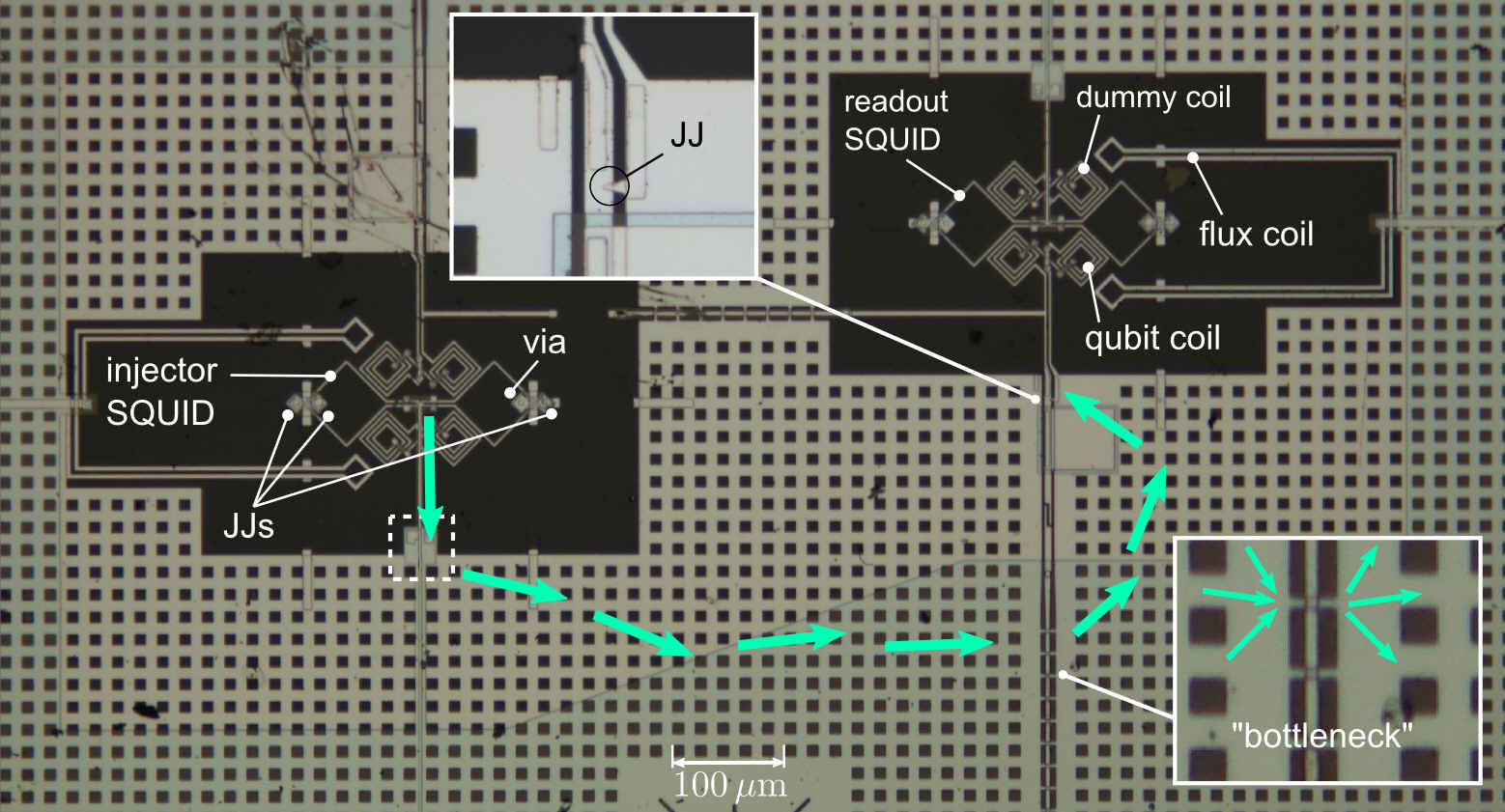}{
		}
		\caption{Photograph of the sample containing two qubits. The QPs are injected by the readout SQUID ("injector SQUID") of the inactive qubit and they diffuse through the ground plane (square perforated gray area) and through the galvanic bridges across the microwave line ("bottleneck") to flow onto the top electrode of the qubit's Josephson junction, which is depicted in the upper inset. The role of the bottleneck is discussed in Appendix~\eqref{sec_QPDiff}.}
		\label{figChip}
	\end{minipage}
\end{figure*}

\begin{figure}[htbp]
	\centering
	\begin{minipage}{0.5\textwidth}
		\begin{flushleft}
%		\psfragfig*[width=0.92\linewidth]{./graphics/pulsesQP/pulsesQP}{
%			\psfrag{aa}{(a)}
%			\psfrag{vinj}{$I_\text{inj}$}
%			\psfrag{tinj}{$\tau_\text{inj}=50-300\,\mathrm{\mu s}$}
%			\psfrag{ttot}{$\tau_{\mathrm{tot}}$}
%			\psfrag{manip}{$\approx 200\,\mathrm{ns}$}
%			\psfrag{op}{qubit manip.}
		\includegraphics[width=0.92\linewidth]{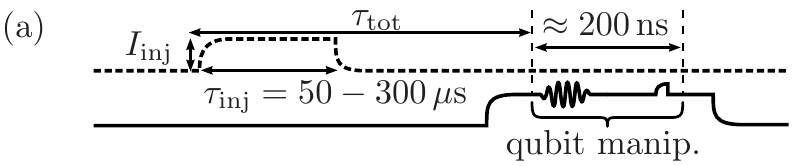}{
		}
		\vspace{0.15cm}
		\end{flushleft}
	\end{minipage}
	\begin{minipage}{0.5\textwidth}
		\vspace{0pt}
		\begin{flushleft}
%		\psfragfig*[width=\linewidth]{./graphics/nqptimedomain/nqptimedomain}
		\includegraphics[width=\linewidth]{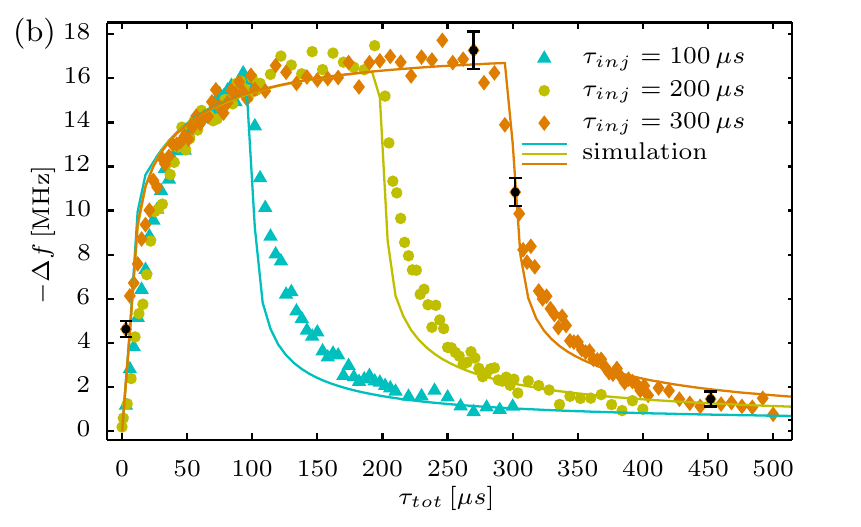}
		\end{flushleft}
	\end{minipage}
	\caption{(a) The pulse arrangement for QP injection. The continuous line indicates the control of the phase qubit and the dashed line indicates the current pulse applied to the injector SQUID. (b) The shift of the qubit transition frequency $-\Delta f$, which linearly scales with the QP density at the JJ~\cite{Lenander2011}, measured as a function of $\tau_\text{inj}$ and for $I_\text{inj}$ which exceeds the switching current $I_{\mathrm{S}}=1.5\,\mathrm{\mu A}$ by a factor of $6$. The black data points indicate the error bars for characteristic regions. The continuous lines are the results of simulations, where the faster decay of $-\Delta f$ for $\tau_{\mathrm{tot}}>\tau_\text{inj}$ is due to the rise time of the injector current pulse. Our measurements on TLS are performed in a regime of stable QP density for $\tau_{\mathrm{tot}}\approx\tau_\text{inj}-50\,\mathrm{\mu s}$.}
	\label{fig_timedomain}
\end{figure}

Figure~\eqref{figChip} shows a photograph of the sample containing the qubit circuit and the injector dc-SQUID, which is galvanically coupled to the JJ via a common thin-film Al ground plane. Similar to the work in Ref.~\cite{Lenander2011}, we apply bias current pulses (of amplitude $I_\text{inj}$) to the injector dc-SQUID exceeding its switching current $I_\text{S}\approx1.5\,\mathrm{\mu A}$ to produce \textit{in situ} QPs from Cooper pair breaking processes, which then diffuse over a distance of $1\,\mathrm{mm}$ through the ground plane towards the JJ. To ensure that $x_{\text{qp}}$ reaches a stationary value during the QP injection, we performed measurements in the time domain. We observed the shift of the qubit resonance frequency $-\Delta f$ that depends linearly on $x_{\text{qp}}$~\cite{Lenander2011} in dependence of varying injection pulse timing. Figure~\eqref{fig_timedomain}(a) illustrates the pulse arrangement used for QP injection, where the continuous line is the flux and microwave control of the phase qubit and the dashed line shows $I_\text{inj}$. In Fig.~\eqref{fig_timedomain}(b), $-\Delta f$ is plotted for several injection pulse widths $\tau_\text{inj}$ vs the time delay  $\tau_{\mathrm{tot}}$ between the start of an injector current pulse of constant amplitude $I_\text{inj}=6.4\,I_{\mathrm{S}}$ and qubit measurement. We see that a stationary QP density is reached for $100\,\mathrm{\mu s}<\tau_{\mathrm{tot}}<\tau_\text{inj}$, i.e., when the injection pulse is sufficiently long and overlaps with the qubit manipulation sequence. In the experiments on TLSs, we therefore inject QPs at $\tau_\text{inj}=200\,\mathrm{\mu s}$ and $\tau_{\mathrm{tot}}=150\,\mathrm{\mu s}$. We have verified the QP diffusion towards the JJ by comparing the data from Fig.~\eqref{fig_timedomain}(b) to a simulation of the QP diffusion process in a simplified 2D chip geometry (Fig.~\eqref{figNQP}). We see a good agreement of simulation data and the measurements, whereas the measured QP density seems to decay slower than predicted by the simulation. This is due to the rise time of the injection pulse.\\

In both independent experiments (the thermal and the injection experiment), we calibrated $x_{\text{qp}}$ by monitoring the response of the qubit as a function of the mixing chamber temperature $T_\text{mch}$ and of the injector current $I_\text{inj}$, respectively. We found that it is favorable to track the qubit's energy-relaxation rate rather than its resonance frequency that is sensitive to quasi static drifts. QPs that tunnel through the JJ can absorb energy from the qubit and lead to qubit decay. We calculate $x_{\text{qp}}$ from the purely QP-induced energy-relaxation rate of the qubit:
\begin{align}
\gamma_1^\text{qub}=&\;\gamma_1^\text{qub,(TB)}+\gamma_1^\text{qub,(BT)},\label{eq_gamma1Full}\\
\gamma_1^\text{qub,(lm)}=&\;\frac{2}{e^2R_\text{T}}t^2\mathit{\Delta}_s\int_{1}^{\infty}d\epsilon\left(1-\frac{\cos{\varphi_0}}{\epsilon(\epsilon+E_\text{q}/\mathit{\Delta}_s)}\right)\notag\\
&\;\times \rho(\epsilon)\rho(\epsilon+\frac{E_\text{q}}{\mathit{\Delta}_s}) f_0^\text{(l)}(\epsilon)\left(1-f_0^\text{(m)}(\epsilon+\frac{E_\text{q}}{\mathit{\Delta}_s})\right)
\label{eq_gamma1}
\end{align}
using the theory by Catelani \textit{et al.}~\cite{Catelani2011}. Here, $\mathit{\Delta}_s$ is the superconducting gap of Al, $t$ is the tunnel element, $R_\text{T}\approx 250\,\mathrm{\Omega}$ is the JJ tunnel resistance, and $e$ is the elementary charge. Both terms in Eq.~\eqref{eq_gamma1Full} result from QP tunneling through the JJ's tunnel barrier from the top electrode to the bottom electrode ("$\text{TB}$") and vice versa ("$\text{BT}$"), respectively. The qubit was tuned to have the eigenenergy $E_\text{q}=h\cdot8.8\,\mathrm{GHz}$ and the mean phase drop across the JJ was $\varphi_0\approx0.4\,\pi$. We approximate the distribution function that depends on the QP temperature $T$ and the chemical potential $\mu$ by the Fermi function $f_0(\epsilon,T,\mu)$ due to the following reasoning: at the injection point, the injected non-equilibrium QPs are expected to show a strong charge imbalance. This so-called charge mode relaxes due to elastic scattering on a time scale of the electron-electron interaction time ($50\,\mathrm{ns}$), which is much smaller than the diffusion time ($100\,\mathrm{\mu s}$) from the injection point to the qubit junction, and by three orders of magnitude smaller than the recombination time of QPs~\cite{Heimes2014}. After the long diffusion path, non equilibrium QPs have thus relaxed into a symmetric distribution very close to $\mathit{\Delta}_s$. Therefore, we use the approximation $f_0(\epsilon,T,\mu)$ to describe the QP distribution in a local equilibrium at the JJ. In the thermal experiment, $\mu=0$ and $T$ is the parameter in Eq.~\eqref{eq_gamma1}, which we adjust via $T_\text{mch}$. In contrast, during the injection experiment, we control $\mu$ via $I_\text{inj}$, whereas $T$ equals the residual QP temperature $T_0$ exceeding the sample temperature, to be discussed in the following paragraph. Due to the fast decay of the charge mode, the polarity of $I_\text{inj}$ does not affect any of the results presented in this work.\\

Without applied injection pulses, we expect to observe an excess QP density $x_\text{qp,0}$ that is higher than its value corresponding to the sample temperature due to QP excitations by infrared photons and from further unknown sources. Shaw \textit{et al.}~\cite{Echternach2008} report about analysis of QP tunneling statistics in charge qubits, from which they deduce $x_\text{qp,0}\approx1.6\times10^{-6}$ at a base temperature of $18\,\mathrm{mK}$. In another experiment by de Visser \textit{et al.}~\cite{Klapwijk2011}, similar QP densities at temperatures below $160\,\mathrm{mK}$ were observed from QP number fluctuations in a superconducting thin-film resonator. We deduce numerically from the common expression for the QP density [Eq. (7)] that the quoted value of $x_\text{qp,0}$ corresponds to a QP temperature $T_0\approx200\,\mathrm{mK}$.\\

Now we explain how we calibrate the QP density in our experiments. At the base temperature of $30\,\mathrm{mK}$ and without injected QPs, the phase qubit relaxes to its ground state at a rate of $\gamma_{1,0}^\text{qub}\approx15\,(\mathrm{\mu s})^{-1}$ owing to interactions with excess QPs and the TLS bath. We obtain the qubit relaxation rate $\gamma_{1}^\text{qub}$ that is solely QP induced by extracting the TLS-induced relaxation rate from the measured qubit's relaxation rate $\gamma_{1,\text{meas}}^{\text{qub}}$: $\gamma_{1}^\text{qub}\equiv\gamma_{1,\text{meas}}^{\text{qub}}-\gamma_{1,0}^\text{qub}+\gamma_1^\text{qub}(T_0)$, where $\gamma_1^\text{qub}(T_0)$ is deduced from Eq.~\eqref{eq_gamma1} and is the small offset in Fig.~\eqref{fig_G1_vs_QP}. There, the resulting $\gamma_{1}^{\text{qub}}$ as a function of both $T_\text{mch}$ and $I_\text{inj}$ is shown, respectively, while $\gamma_{1,\text{meas}}^{\text{qub}}$ was recorded at timing parameters $\tau_{\mathrm{tot}}=150\,\mathrm{\mu s}$ and $\tau_{\mathrm{inj}}=200\,\mathrm{\mu s}$. We numerically deduce $T$ or $\mu$ from $\gamma_{1}^\text{qub}$ in the thermal or injection experiment, respectively. Then we calculate $x_\text{qp}$ (see right vertical axis). The corresponding fits (continuous lines) provide the calibration of $x_\text{qp}$ vs $I_\text{inj}$ and $T_\text{mch}$, respectively.\\

%--------------------SECTION----------------------------------------
\section{Imbalance of QP densities in the injection experiment}
\label{sec_QPImbalance}
In Fig.~\eqref{figChip}, we show that the injected QPs appear in the top electrode of the qubit's JJ (see top inset of the photograph). From that point, QPs diffuse either through the qubit's coil that is about $750\,\mathrm{\mu m}$ long or they tunnel through the JJ onto the bottom electrode. Due to this detour and due to relatively low tunnel rates through the JJ, it is possible that the stationary QP densities on both electrodes may show an imbalance. We have solved the stationary Boltzmann equation and found the imbalance $\alpha=x_\text{qp}^\text{(T)}/x_\text{qp}^\text{(B)}$ to be around 4 when assuming no tunneling and 2 for typical tunneling rates of $\approx 6\,(\mathrm{\mu s})^{-1}$. The measured QP density $x_\text{qp}$ is deduced numerically from the detected qubit's energy-relaxation rate shown in Eq.~\eqref{eq_gamma1Full}. By simplifying the integral in Eq.~\eqref{eq_gamma1}, one finds an analytical solution that gives satisfying results:
\begin{align}
\gamma_1^\text{qub}\propto \frac{1}{2} (x_\text{qp}^\text{(B)}+x_\text{qp}^\text{(T)}).\label{eq_gamma1Xqp}
\end{align}
Thus, $x_\text{qp}$ is the average of the QP densities on both electrodes:
\begin{align}
x_\text{qp}=\frac{1}{2}(x_\text{qp}^\text{(B)}+x_\text{qp}^\text{(T)}).\label{eq_xqpMeas}
\end{align}
When we generate QPs by increasing the sample temperature, QPs appear evenly on both sides of the JJ, and accordingly, $x_\text{qp}=x_\text{qp}^\text{(T)}=x_\text{qp}^\text{(B)}$. In contrast, when injecting QPs, the measured QP density is $x_\text{qp}=(x_\text{qp}^\text{(B)}+\alpha x_\text{qp}^\text{(B)})/2$. We thus can deduce from the measured value of $x_\text{qp}$ and an assumed value for $\alpha$ the corresponding QP densities in the electrodes:
\begin{align}
x_\text{qp}^\text{(B)}=&\; x_\text{qp}\frac{2}{(1+\alpha)},\notag\\
x_\text{qp}^\text{(T)}=&\; x_\text{qp}\frac{2\alpha}{(1+\alpha)}.
\label{eq_xqpImbalance}
\end{align}
 To cross check this calculation, we set $\alpha=1$ and get the same results as for the thermal experiment.

%--------------------SECTION----------------------------------------
\section{Heating of the sample by quasiparticle injection?}
\label{sec_QPHeating}
We inject QPs by driving the Josephson junctions (JJs) of the injector SQUID into their resistive state where heating may occur. Moreover, the injected QPs relax by recombination and by inelastic scattering on phonons and impurities. Those processes result in phonon creation which can lead to heating. To inspect the sample temperature, we have used the readout dc-SQUID as a sensitive thermometer, which is placed close to the qubit coil (see. Fig. \eqref{figChip}, top right corner) and at a linear distance of about $500\,\mathrm{\mu m}$ from the injector SQUID. The switching current of a JJ decreases linearly with increasing sample temperature once the thermal activation rate exceeds the tunneling rate. The associated threshold temperature is called the cross over temperature~\cite{GOW1987}. Also, the standard deviation $\sigma$ of the ensemble of switching currents acquired in the current-ramp measurement increases linearly with temperature above the cross over temperature~\cite{Wallraff03}. Properties of the phase qubit used in this work such as its energy-relaxation rate $\gamma_1^\text{qub}$ change significantly for sample temperatures exceeding $200\mathrm{mK}$, whereas the cross over temperature of the readout SQUID is less than $30\,\mathrm{mK}$, making it a much more sensitive detector for the sample temperature than the qubit. We have measured the increase $\Delta\sigma$ of the switching-current standard deviation as a function of the cryostat's mixing-chamber temperature $T_\text{mch}$ and as function of QP injector current $I_\text{inj}$, respectively, to compare both behaviors.\\

\begin{figure}[htbp]
	\begin{minipage}{0.5\textwidth}
		\begin{flushleft}
			\includegraphics[width=0.78\linewidth]{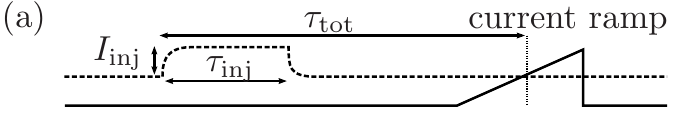}
%			}
		\end{flushleft}
	\end{minipage}
	\begin{minipage}{0.5\textwidth}
		\begin{flushleft}
%		  	\psfragfig*[width=0.81\linewidth]{graphics/stddevV/stddevV}{
%			  	\psfrag{(b)}{(b)}
%				\psfrag{td}{\small$\tau_\text{tot}$}
%				\psfrag{sig}[c][c][1][90]{$\mathrm{\Delta\sigma\;[A]}$}
%				\psfrag{v}{$\mathrm{|I_{inj}|\;[\mu A]}$}
		  	\includegraphics[width=0.81\linewidth]{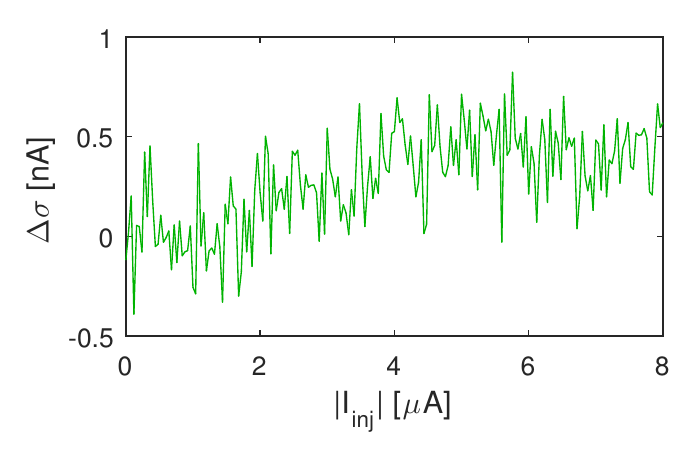}{
		  	}
		\end{flushleft}
	\end{minipage}
	\centering
	\begin{minipage}{0.5\textwidth}
		\begin{flushleft}
%			\psfragfig*[width=0.81\linewidth]{graphics/stddevT/stddevT}{
%			  	\psfrag{(c)}{(c)}
%				\psfrag{l}{$\ket{L}$}
%				\psfrag{r}{$\ket{R}$}
%				\psfrag{ds}[c][c][1][90]{$\mathrm{\Delta\sigma\;[A]}$}
%				\psfrag{tk}{$\mathrm{T_{mch}\;[K]}$}
			\includegraphics[width=0.81\linewidth]{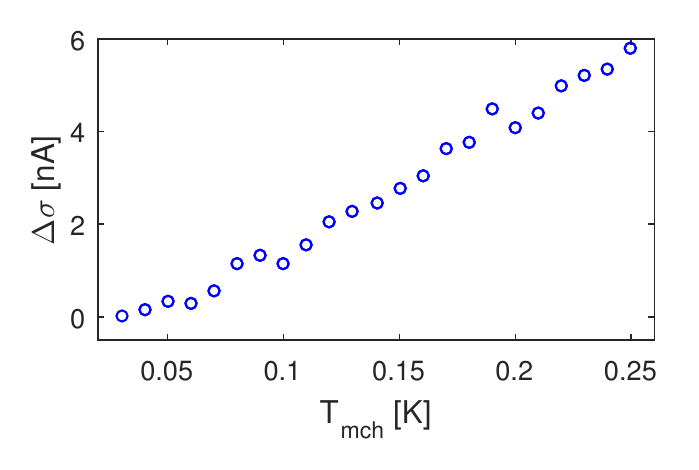}{
			}
		\end{flushleft}
	\end{minipage}
	\caption{(a) Pulse sequence to record the switching-current statistics vs the QP injector current $I_\text{inj}$. (b) $\Delta\sigma$ vs $I_\text{inj}$. $\Delta\sigma$ is the increase of the standard deviation in the measurement of the SQUID's switching current. Here, $\tau_{\mathrm{inj}}\approx200\,\mathrm{\mu s}$ and $\tau_{\mathrm{tot}}\approx225\,\mathrm{\mu s}$. (c) $\Delta\sigma$ in dependence of the mixing-chamber temperature. The linear increase is clear. By comparing the maximum $\Delta\sigma$ in (b) and (c), we conclude that during QP injection, heating is negligible.}
	\label{figSQUIDprobing}
\end{figure}

Figure \eqref{figSQUIDprobing}(a) shows the pulse arrangement to measure $\Delta\sigma$ vs $I_{\mathrm{inj}}$. In Fig. \eqref{figSQUIDprobing}(b), we see the acquired data of $\Delta\sigma$ when injecting QPs. The injection pulse width was $\tau_{\mathrm{inj}}\approx200\,\mathrm{\mu s}$ and $\tau_{\mathrm{tot}}\approx225\,\mathrm{\mu s}$, which in this experiment is the delay between the beginning of the injection pulse and the middle of the current ramp (the ramp is ca. $200\,\mathrm{\mu s}$ wide). The qubit is not operated in this experiment, and accordingly no microwave tones are applied. We read an average broadening from zero to maximal injection of about $\Delta\sigma\approx0.4 \,\mathrm{nA}$.\\

The temperature-related $\Delta \sigma$ was also measured when $I_\text{inj}$ was zero. In Fig. \eqref{figSQUIDprobing}(c), we present $\Delta\sigma$ vs $T_\text{mch}$ that was varied from $30\,\mathrm{mK}$ to $250\,\mathrm{mK}$. An immediate increase of $\Delta\sigma$ confirms that the SQUID's cross over temperature is below $30\,\mathrm{mK}$. The increase of $\Delta\sigma$ is about $25\,\mathrm{nA/K}$, whereas during QP injection, $\Delta\sigma$ remains below $0.4\,\mathrm{nA}$, corresponding to a temperature of $45\,\mathrm{mK}$. This temperature is negligible as compared to $200\,\mathrm{mK}$, beyond which the qubit's energy-relaxation increases significantly (see Fig. ~\eqref{fig_G1_vs_QP}). Hence, the SQUID-mediated injection of QPs works reliably, controllably, and mostly free of undesired heating. This is an important finding for our experiments on QP-induced decoherence of TLSs because, at low temperatures, the simplest explanation of any coherence-breaking effect, when ohmic currents are applied, would be heating.\\

%--------------------SECTION----------------------------------------
\section{Simulation of the diffusion of quasiparticles}
\label{sec_QPDiff}
Here we discuss the simulations we performed to understand the diffusion process of injected QPs towards the qubit's Josephson junction (JJ). Rothwarf and Taylor~\cite{RothwarfTaylor1076} showed that during the thermalization of QPs in thin superconducting films, the phonons created from a QP recombination event have a high probability to be involved in a new Cooper pair breaking process before they relax to the thermal level. This so-called phonon trapping motivates one to consider the QPs and the non thermal phonons as two coupled fluids. Here we introduce the Rothwarf-Taylor equations (RT equations), add diffusive terms, and discuss why we may uncouple the RT equations and uniquely regard the QP diffusion equation in our simulations.\\

\begin{figure*}[htbp]
	\begin{minipage}{0.49\textwidth}
%  	\psfragfig*[width=\linewidth]{./graphics/nQPRising/nQPRising}{
%	  	\psfrag{a}{(a)}
%  		\psfrag{nqp}{$x_{\mathrm{qp}}$}
%  		\psfrag{V}{$V_\mathrm{inj}=1\,\mathrm{V}$}
%  		\psfrag{inj}{$\tau_\mathrm{inj}=400\,\mathrm{\mu s}$}	
%  		\psfrag{dif}{$\tau_\mathrm{tot}=14\,\mathrm{\mu s}$}	
%  		\psfrag{bottle}{bottleneck}
%  		\psfrag{200um}{$200\,\mathrm{\mu m}$}
  	\includegraphics[width=\linewidth]{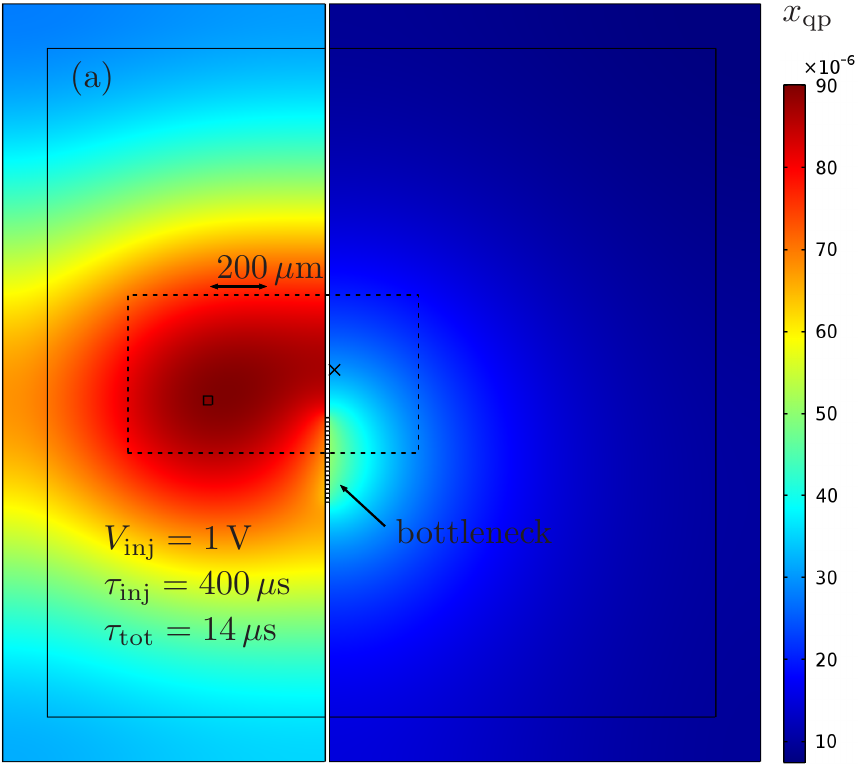}{
  	}
	\end{minipage}
	\begin{minipage}{0.49\textwidth}
%	\psfragfig*[width=\linewidth]{./graphics/nQPStationary/nQPStationary}{
%	  	\psfrag{b}{(b)}	
%		\psfrag{nqp}{$x_{\mathrm{qp}}$}
%		\psfrag{V}{$V_\mathrm{inj}=1\,\mathrm{V}$}
%		\psfrag{inj}{$\tau_\mathrm{inj}=400\,\mathrm{\mu s}$}	
%		\psfrag{dif}{$\tau_\mathrm{tot}=300\,\mathrm{\mu s}$}	
%		\psfrag{bottle}{bottleneck}
%		\psfrag{200um}{$200\,\mathrm{\mu m}$}
	\includegraphics[width=\linewidth]{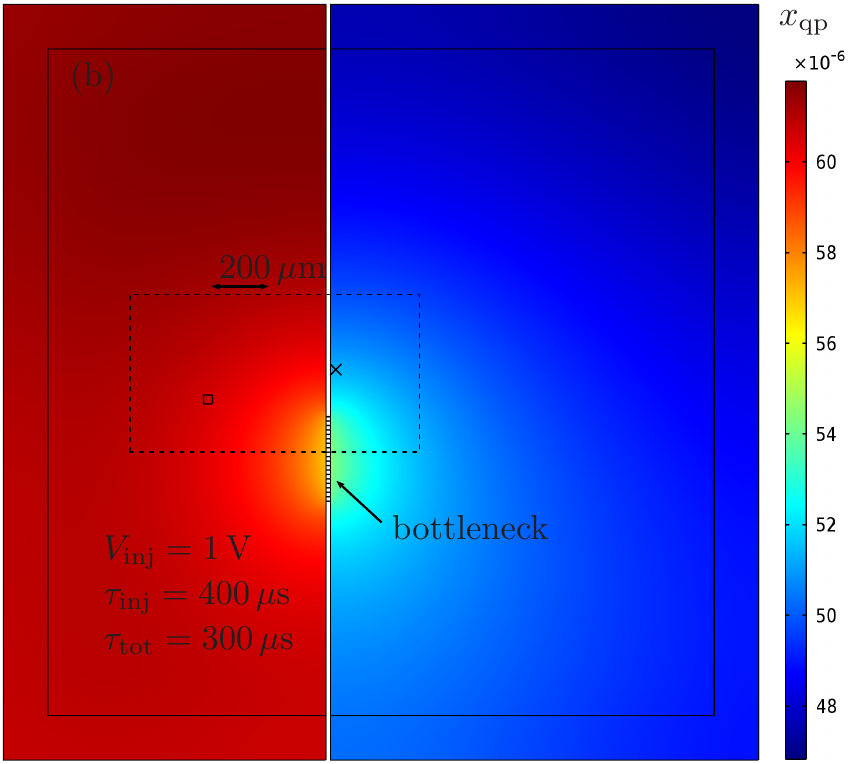}{
	}
	\end{minipage}
	\caption{Simulated quasiparticle density $x_{\mathrm{qp}}$ in the simplified 2D geometry for $I_{\mathrm{inj}}=9.8\,\mathrm{\mu A}$ and $\tau_\mathrm{inj}=400\,\mathrm{\mu s}$. The dashed rectangle indicates the size of the chip photograph in Fig. \eqref{figChip}. The tiny rectangle denotes the effective QP injection point and the cross denotes the QP destination site, i.e., the qubit's JJ. The twenty $10\times2\,\mathrm{\mu m}$ bridges reconstruct the bottleneck in the real geometry. (a) $x_{\mathrm{qp}}$ in the non stationary injection regime for $\tau_\mathrm{tot}=14\,\mathrm{\mu s}$, which is the delay between the start of injection and measurement [see Fig.~\eqref{fig_timedomain}(a)]. (b) The stationary regime for $\tau_\mathrm{tot}=300\,\mathrm{\mu s}$. Here, we clearly recognize the bottleneck reducing the stationary maximum QP density by about $12\%$.}
	\label{figNQP}
\end{figure*}

The detailed derivation of the RT equations is shown in Ref.~\cite{ChangScalapino1977}. Due to phonon trapping, we have to consider both the QP density $n_\text{qp}$ and the phonon density $N_\text{ph}$ whose time dependencies are coupled. The phonons contributing to QP generation have an energy surpassing $2\mathit{\Delta}_s$, which we now call "hot" phonons with a given density,
\begin{align}
N_{\mathrm{ph}}=\mathit{\Delta}_s\int_{2}^\infty d\Omega D_{\mathrm{ph}}(\Omega) g(\Omega),
\end{align}
where $\Omega$ is the phonon energy divided by $\mathit{\Delta}_s$ and $D_{\mathrm{ph}}(\Omega)$ and $g(\Omega)$ are the phonon density of states and distribution function, respectively. Now both quantities can be related by the RT equations~\cite{RothwarfTaylor1076}:
\begin{align}
\dot n_{\mathrm{qp}}=&-2Rn_{\mathrm{qp}}^2+2BN_{\mathrm{ph}}+I_{\mathrm{qp}},\label{eqNQP}\\
\dot N_{\mathrm{ph}}=&+\phantom{2}Rn_{\mathrm{qp}}^2-\phantom{2}BN_{\mathrm{ph}}-\frac{N_{\mathrm{ph}}-N_0}{\tau_{\mathrm{esc}}}.
\label{eqNPH}
\end{align}
Here, $R$ denotes the QP recombination constant in units of $[m^3/s]$ and $B$ is the QP recreation rate from phonon trapping. The factor 2 in the exchange terms in Eq. \eqref{eqNQP} designates that the QP recombination and creation process always involves two QPs and one phonon. $I_{\mathrm{qp}}$ is the injection current density of QPs. The last term in Eq. \eqref{eqNPH} accounts for phonon escape into the substrate, $-N_{\mathrm{ph}}/\tau_{\mathrm{esc}}$, and the return of phonons from the substrate, $+N_0/\tau_{\mathrm{esc}}$, where $\tau_{\mathrm{esc}}$ represents the escape time. The substrate is supposed to be in thermal equilibrium due to its much larger volume compared to the thin-film, thus the return term is constant and can even be neglected, as the thermal contribution of the substrate to "hot" phonons is negligible at our sample temperature of $30\,\mathrm{mK}$. Further, at such low temperature, the phonon-phonon scattering that scales with $T^4$ can be neglected so that phonons of energies $\Omega\geq2\mathit{\Delta}_s$ can be assumed to move nearly ballistically in the superconducting film. The (slower) transversal phonons propagate at a velocity of $v=3050\,\mathrm{m/s}$ in Al. Considering that we inject QPs at a maximal energy of $6.4\mathit{\Delta}_s$, we estimate the mean diffusion constant of QPs to be $D\approx 22.5\,\mathrm{cm^2/s}$~\cite{Martinis2009}. Now, the propagation time scales of the phonons and QPs can be compared. The rule of thumb for diffusion states: the diffusing particle covers a distance of $\sqrt{D\tau}$ in time $\tau$, whereas a phonon would need the time $\sqrt{D\tau}/v$ for the same distance. Thus, for a distance of, e.g., $100\,\mathrm{\mu m}$, the QP would need approximately $4\,\mathrm{\mu s}$ and a phonon $0.03\,\mathrm{\mu s}$. Hence, phonons move much faster in the superconducting film than QPs; consequently they react almost instantaneously to any change in the QP ensemble and they can be considered in the stationary regime. We thus may set Eq.~\eqref{eqNPH} to zero and get
\begin{align}
N_{\mathrm{ph}}=&\frac{Rn_{\mathrm{qp}}^2}{B+1/\tau_{\mathrm{esc}}},\notag\\
\dot n_{\mathrm{qp}}=&-n_{\mathrm{qp}}^2 2R\left(1-\frac{B}{B+1/\tau_{\mathrm{esc}}}\right) + I_{\mathrm{qp}},\notag\\
\equiv&-\widetilde Rn_{\mathrm{qp}}^2 + I_{\mathrm{qp}}.
\label{eqNRed}
\end{align}
Here we decoupled the RT equations and reduced them to the single QP decay equation [Eq.~\eqref{eqNRed}] with constant injection, where we have defined the effective recombination constant $\widetilde R$. Further, we have to adapt the decay equation \eqref{eqNRed} to our experiment, where the injection point is distant from the measuring point. We thus add a diffusion term including the second spatial derivative of the QP density $\boldsymbol{\nabla}^2n_{\mathrm{qp}}$ weighted with the homogeneous diffusion constant $D$:
\begin{align}
\dot n_{\mathrm{qp}}-D\boldsymbol{\nabla}^2n_{\mathrm{qp}}=&-\widetilde R n_{\mathrm{qp}}^2 + I_{\mathrm{qp}}.
\label{eqDiff}
\end{align}
Equation~\eqref{eqDiff} is the final diffusion equation which was used to simulate the space- and time-dependent QP density $x_{\mathrm{qp}}\equiv n_{\mathrm{qp}}/n_{\mathrm{cp}}$ with the Comsol software package\cite{Comsol}, where $n_{\mathrm{cp}}$ is the constant density of Cooper pairs. In Fig. \eqref{figChip}, we see the photograph of the chip, where the bold green arrows show the shortest path ($\approx1\,\mathrm{mm}$) for the QPs to diffuse from the injection point (injector SQUID) to the qubit's JJ. The most important feature of the thin-film layout is the conducting bridges spanning the microwave line that pose a bottleneck for the diffusing QPs. They are reconstructed in the simplified 2D simulation geometry by twenty $10\,\mathrm{\mu m}\times2\,\mathrm{\mu m}$ strips that connect both parts of the ground plane (see Fig.~\eqref{figNQP}). The squared holes all over the aluminium film contribute to an effective QP constant of diffusion; thus they are not considered in the simulation geometry. Another detail is the ca. $140\,\mathrm{\mu m}\times 2\,\mathrm{\mu m}$ large strip leading from the Josephson junctions of the injector SQUID to the ground plane. Here QPs are more confined and are expected to recombine faster, but as this constriction applies at the very beginning of the diffusion path, this gives only an effective, reduced injection current density $\widetilde I_{\mathrm{qp}}$. For this reason, in the simulation, the effective injection point has been chosen to be the contact point of the strip to the ground plane (white dashed square in Fig \eqref{figChip}).\\

In Fig. \eqref{figNQP}, we see the simulation data for $I_{\mathrm{inj}}=6.4\,I_{\mathrm{S}}$, while the parameters used in the simulation are shown in the Table~\eqref{tabQP}. The color of the surface plots denotes the normalized QP density $x_{\mathrm{qp}}$ within the simulation geometry. The dashed rectangle shows the size of the sample photograph in Fig. \eqref{figChip}, the cross is the qubit's JJ, and the tiny black square indicates the effective injection point. There we recognize the bottleneck connecting both sides of the aluminium ground plane, which reduces the stationary maximum QP density by about $12\%$ on the side of the ground plane connected to the JJ. Figure~\eqref{figNQP}(a) shows the QP distribution shortly after the start of the injection ($\tau_\mathrm{tot}=14\,\mathrm{\mu s}$, $\tau_\mathrm{inj}=400\,\mathrm{\mu s}$; see Fig. 2 in the main text). In Fig.~\eqref{figNQP}(b) the stationary case for $\tau_\mathrm{tot}=300\,\mathrm{\mu s}$ ($\tau_\mathrm{inj}=400\,\mathrm{\mu s}$) is shown. In the area between the simulation geometry border and the inner rectangle (continuous black line), additional linear QP relaxation was added to avoid boundary effects such as QP reflection. This area shall effectively enlarge the simulation geometry in order to minimize the meshing grid and the calculation time. $x_{\mathrm{qp}}$ has been simulated as a function of $\tau_{\mathrm{tot}}$ for various $\tau_\mathrm{inj}$ and for some injection amplitudes. Subsequently, $x_\text{qp}$ has been transferred into the frequency shift of the qubit $\mathit{\Delta} f$ ($\mathit{\Delta} f$ is proportional to $x_\text{qp}$~\cite{Catelani2011}) to compare it with the measured $\mathit{\Delta} f$ in a QP injection experiment, as shown in Fig.~\eqref{fig_timedomain}.\\
\begin{table}
\centering
\begin{tabular}{ |c|c|c| }
%\centering
\hline
 & Simulation & Literature\\ \hline
Diffusion constant $D\,(\mathrm{cm^2/s})$ & $22.5$ & $ 22.5$~\cite{Martinis2009} \\ \hline
Recombination const. $\widetilde R\,(\mathrm {m^3/s})$ & $1.5\cdot 10^{-17}$ &  $1.5\cdot 10^{-17}$~\cite{Heimes2014} \\ \hline
Injection current dens. $\widetilde I\,(\mathrm {1/m^3s})$ & $8.5\cdot 10^{30}$ & -\\ \hline
\end{tabular}
\caption{Parameters used to simulate the diffusion of quasiparticles from the injection point to the Josephson junction.}
\label{tabQP}
\end{table}

%--------------------SECTION----------------------------------------
\section{QP tunneling through a Josephson junction}
\label{sec_QPTunnel}
In this section we estimate the penetration depth of the evanescent QP wave function in the $\mathrm{AlO_x}$ tunnel barrier of the JJ. We need this quantity to discuss the coupling strength of QPs to TLSs in dependence of the TLS' position across the JJ.\\

We model the QP tunneling through the JJ by a plane wave of energy $\approx E_F$ that tunnels through a 1D rectangular potential wall of an unknown height, $V_0>E_F$. The spatial coordinate $x$ is taken along the normal vector to the surface of the JJ electrode, whereas the wall spans the distance from $x=0$ to $x=d\approx 2\,\mathrm{nm}$ ($d$ is the thickness of the tunnel barrier). The solution within the wall decays exponentially on a spatial scale of $\rho^{-1}=(2m(V_0-E_F)/\hbar)^{-1/2}$. The transmission coefficient $T$ for the incident wave through the potential wall is given by
\begin{equation}
T=\frac{4E_F(V_0-E_F)}{4E_F(V_0-E_F)-V_0^2\sinh^2(\rho d)}.
\label{eq_transmission}
\end{equation}
%The Al/AlOx/Al Josephson junction of the qubit used in this work has a tunnel resistance of about $250\,\mathrm{\Omega}$. From this we estimate
The typical QP tunneling rate through the JJ is $\approx 6\,(\mathrm{\mu s})^{-1}$, which is the product of its attempt rate $E_F/h=3\times10^9\,\mathrm{(\mu s)}^{-1}$ and the transmission coefficient $T$. From this, we get $T\approx2\times10^{-9}$ and we deduce numerically from Eq.~\eqref{eq_transmission} $V_0\approx 13.3\,\mathrm{eV}$, whereas $E_F=11.7\,\mathrm{eV}$ for aluminun. The effective electron mass in aluminun is $1.1$ times the electron mass $m_e$ so that the penetration depth of QPs within the tunnel barrier turns out to be $\rho^{-1}\approx 0.15\,\mathrm{nm}$.\\
%--------------------SECTION----------------------------------------
\section{Interaction of TLS with QP and estimated TLS position across the tunnel barrier}
\label{sec_TLSPosition}
\begin{figure}[htbp]
	\begin{minipage}{0.49\textwidth}
%		\psfragfig*[width=\linewidth]{./graphics/fittlsnqp_loglog/fittlsnqp_loglog}
		\includegraphics[width=\linewidth]{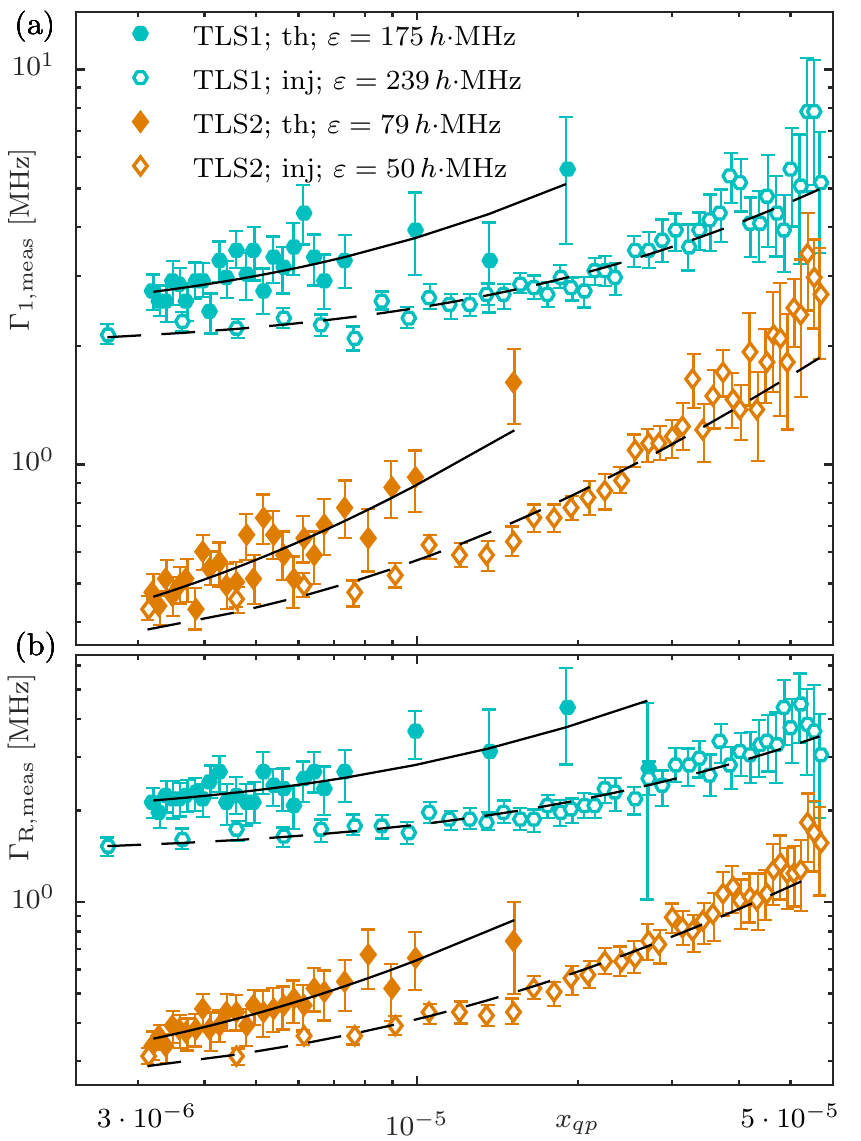}		
	\end{minipage}
	\caption{These measurement data and fits are presented in the main text in Fig.~\eqref{fig_fittlsnqp}. However, this double-logarithmic plot is more readable for low  $x_\text{qp}$, whereas the linear fits are not obvious like in Fig.~\eqref{fig_fittlsnqp}. (a) Measured energy-relaxation rates $\Gamma_{1,\text{meas}}$ of TLS1 and TLS2. The legends indicate the asymmetry energy $\varepsilon$ and whether quasiparticles were injected or thermally generated. Black lines are theoretical fits to Eq. (6). (b) Recorded decay rates $\Gamma_\text{R,meas}$ of TLS Rabi oscillations vs $x_\text{qp}$ and the corresponding fits.}
	\label{fig_fittlsnqp_loglog}
\end{figure}

\begin{figure}[htbp]
	\begin{minipage}{0.49\textwidth}
%  	\psfragfig*[width=\linewidth]{./graphics/Kratio/Kratio}{
%	  	\psfrag{rho}{$\rho$}
  	\includegraphics[width=\linewidth]{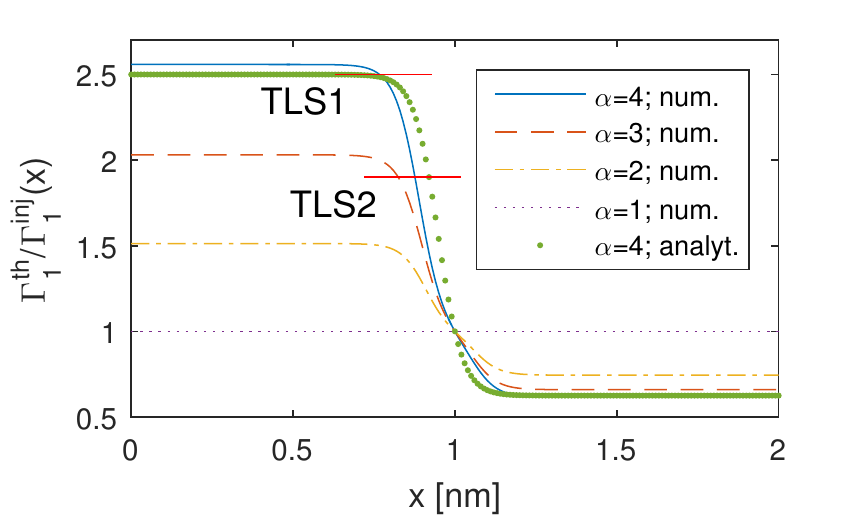}{
  	}
	\end{minipage}	
	\caption{The ratio $\Gamma_1^\text{th}/\Gamma_1^\text{inj}(x)$ as a function of the TLS position across the $2-\,\mathrm{nm}$-thick tunnel barrier, where $x$ equals zero at the bottom electrode. $\Gamma_1^\text{th}(x)$ is the theoretical prediction for the TLS' energy-relaxation rate when QPs are thermally generated and $\Gamma_1^\text{inj}(x)$ is valid when injecting QPs. The QP imbalance $\alpha$ is indicated in the legend. "num." designates that the ratio was numerically calculated from Eq.~\eqref{eq_gamma1ofx}, whereas the graph labeled as "analyt." shows the approximation from Eqs.~\eqref{eq_gamma1QPTherm} and~\eqref{eq_gamma1QPInject}. The ratio $\Gamma_1^\text{th}/\Gamma_1^\text{inj}$ equals the ratio $K^\text{th}/K^\text{inj}$ (see red horizontal lines labeled by TLS1 or TLS2) of the fit factors presented in the main text, from which one can estimate the positions of TLS1 and TLS2 to be roughly in the middle but closer to the bottom electrode, and $\alpha$ to be approximately $4$.}
	\label{fig_Kratio}
\end{figure}

\begin{figure}[htbp]
	\begin{minipage}{0.49\textwidth}
%  	\psfragfig*[width=\linewidth]{./graphics/K1/K1}{
%	  	\psfrag{aa}{(a)}
  	\includegraphics[width=\linewidth]{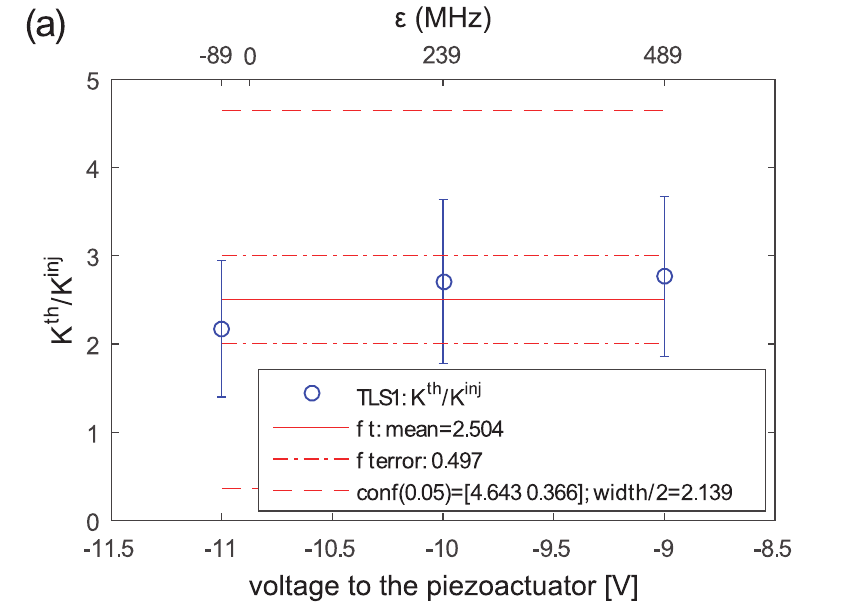}{
  	}
	\end{minipage}
	\begin{minipage}{0.49\textwidth}
%  	\psfragfig*[width=\linewidth]{./graphics/K2/K2}{
%	  	\psfrag{aa}{(b)}
  	\includegraphics[width=\linewidth]{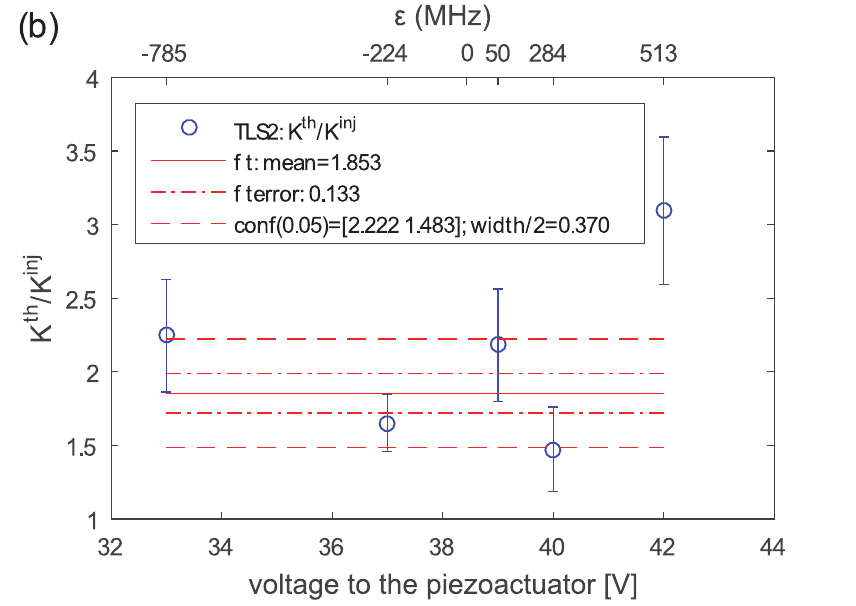}{
  	}
	\end{minipage}	
	\caption{The fit factor ratio $K^\text{th}/K^\text{inj}$ for varying values of the voltage $\mathrm{V_p}$ applied to the piezoactuator (bottom axes)[see Appendix~\eqref{sec_piezo}] and for two probed TLSs: (a) TLS1 and (b) TLS2. The corresponding asymmetry energies $\varepsilon$ are indicated on the non linear top axes. $K^\text{th}$ and $K^\text{inj}$ are the fit factors explained in the main text. The continuous red line is the mean value, the dot-dashed line designates the one sigma interval around the mean value, and the dashed line indicates the $5\%$ confidence interval.}
	\label{fig_Kratio_meas}
\end{figure}

In this section, we offer an explanation for our observation on the TLS' response to quasiparticles: when thermally generating quasiparticles, the TLS' decoherence rate is about twice as high as in the case of injected quasiparticles. In Fig.~\eqref{fig_fittlsnqp} and in Fig.~\eqref{fig_fittlsnqp_loglog}(which is more readable for small values of $x_\text{qp}$), one can see this discrepancy when comparing the decoherence rates at any given value of $x_\text{qp}$.\\
As mentioned in the main text, at temperatures $T<E/\text{k}_\text{B}$, the TLS' energy-relaxation rate $\Gamma_1$ increases with the QP density, whereas the contribution by phonons remains constant. The scattering Hamiltonian in Eq. (4) takes into account only QPs that return into the initial electrode after scattering on a TLS. The full Hamiltonian has the form
\begin{align}
\widetilde{H}_\text{QP}=\;&2g(\frac{\mathit{\Delta}}{E}\tau_x + \frac{\varepsilon}{E}\tau_z)\nonumber\\
&\times\sum_{k,k',l,m}(e^{i\varphi\varepsilon_{lm3}/2}u^{(l)}_{k}u^{(m)}_{k'}-e^{-i\varphi\varepsilon_{lm3}/2}v^{(m)}_{k'}v^{(l)}_{k})\nonumber\\
&\times\sqrt{s_{l}}\alpha^{(l)\dagger}_{k}\sqrt{s_{m}}\alpha^{(m)}_{k'},
\label{eq_HamQPTLSJJ_Full}
\end{align}\\
where we sum over the top and the bottom electrodes ($l,m\in\{B,T\}$). The epsilon tensor $\varepsilon_{l,m,3}$ (for which, without loss of generality, $\{B,T\}\equiv\{1,2\}$) implies that when a QP is backscattered into the initial electrode of the JJ ($l\text{=}m$), it does not couple to the phase drop $\varphi$ across the JJ. The position $x\in[0..d]$ of the probed TLS across the tunnel barrier is contained in the prefactors $(s_{(l)})^{1/2}$ of the QP wave functions that implicate their exponential decay,
\begin{equation}
s_\textbf{B}(x)=e^{-2\rho x};\;s_\textbf{T}(x)=e^{-2\rho(d-x)},
\label{eq_sparms}
\end{equation}
where $d=2\,\mathrm{nm}$ is the tunnel barrier's thickness and the position $x=0$ is at the bottom electrode. The penetration depth of QPs into the tunnel barrier, $\rho^{-1}\approx 0.15\,\mathrm{nm}$, has been estimated in Appendix~\eqref{sec_QPTunnel}. Using Fermi's golden rule, the energy-relaxation rate of the probed TLS reads
\begin{align}
\Gamma_1(x)=&\;s_\textbf{B}^2\Gamma_{1}^{\text{(B)}} + s_\textbf{T}^2\Gamma_{1}^{\text{(T)}} + s_\textbf{B}^{ }s_\textbf{T}^{ }\left(\Gamma_{1}^{\text{(BT)}} + \Gamma_{1}^{\text{(TB)}}\right), \label{eq_gamma1ofx}\\
\Gamma_1^{(lm)}=&\;\frac{4\pi}{\hbar}(N_0Vg\frac{\mathit{\Delta}}{E})^2\mathit{\Delta}_s\int_{1}^{\infty}d\epsilon\left(1-\frac{\cos(\varphi_0)}{\epsilon(\epsilon+E/\mathit{\Delta}_s)}\right)\notag\\
&\times \rho(\epsilon)\rho(\epsilon+\frac{E}{\mathit{\Delta}_s}) f_0^{(l)}(\epsilon)\left(1-f_0^{(m)}(\epsilon+\frac{E}{\mathit{\Delta}_s})\right).\label{eq_gamma1_TB}
\end{align}
where the position dependence is contained in $s_{(l)}(x)$. The first two terms in Eq.~\eqref{eq_gamma1ofx} stand for backscattered QPs into the initial electrode and the right term represents the scattering from the bottom into the top electrode, and vice versa. $s_\textbf{B}^{ }s_\textbf{T}^{ }=\exp\{-2\rho d\}$ is a small value; for this reason, it was neglected in the main text for better readability. $\Gamma_{1}^{\text{(l)}}$ is defined in Eq.~\eqref{eq_gamma1TLS_B} and $\varphi_0\approx0.4\,\pi$ is the mean phase drop across the JJ. Both, $\Gamma_{1}^{\text{(l)}}$ and $\Gamma_{1}^{\text{(lm)}}$ are approximately proportional to $x_\text{qp}^{(l)}$; thus we simplify:
\begin{equation}
\Gamma_1(x)\propto s_\textbf{B}^2x_\text{qp}^{\text{(B)}} + s_\textbf{T}^2x_\text{qp}^{\text{(T)}} + s_\textbf{B}^{ }s_\textbf{T}^{ }\left(x_\text{qp}^{\text{(B)}} + x_\text{qp}^{\text{(T)}}\right). \label{eq_gamma1QP}
\end{equation}
In the thermal experiment, when increasing the temperature $T_\text{mch}$, we thermally generate the same QP density on both electrodes. Thus, $x_\text{qp}=x_\text{qp}^\text{(B)}=x_\text{qp}^\text{(T)}$ and the TLS' energy-relaxation rate induced by thermally generated QPs reads
\begin{equation}
\Gamma_1^\text{th}(x)\propto x_\text{qp}\left(s_\textbf{B}^2 + s_\textbf{T}^2 + 2s_\textbf{T}^{ }s_\textbf{B}^{ }\right). \label{eq_gamma1QPTherm}
\end{equation}
However: in the injection experiment, when injecting QPs, the imbalance $\alpha$ has to be taken into account (see Appendix~\eqref{sec_QPImbalance}):
\begin{equation}
\Gamma_1^\text{inj}(x)\propto x_\text{qp}\left(s_\textbf{B}^2\frac{2}{1+\alpha} + s_\textbf{T}^2\frac{2\alpha}{1+\alpha} + 2s_\textbf{T}^{ }s_\textbf{B}^{ }\right). \label{eq_gamma1QPInject}
\end{equation}
In Fig.~\eqref{fig_Kratio}, we present the ratio $\Gamma_1^\text{th}/\Gamma_1^\text{inj}$ as a function of $x$. The legend designates the $\alpha$ value and whether a numerical calculation using Eq.~\eqref{eq_gamma1ofx} was performed ("num.") or the approximation from Eqs.~\eqref{eq_gamma1QPTherm} and~\eqref{eq_gamma1QPInject} was used ("analyt."). The ratio $K^\text{th}/K^\text{inj}$ of the fit factors presented in the main text [Fig.~\eqref{fig_fittlsnqp}] corresponds to the ratio $\Gamma_1^\text{th}/\Gamma_1^\text{inj}$. In Fig.~\eqref{fig_Kratio_meas}, the $K^\text{th}/K^\text{inj}$ ratios for TLS1 [Fig.~\eqref{fig_Kratio_meas}(a)] and TLS2 [Fig.~\eqref{fig_Kratio_meas}(b)] are plotted vs the voltage $\mathrm{V_p}$ applied to the piezoactuator that changes the TLS asymmetry energy $\varepsilon$ of the TLS [see Appendix.~\eqref{sec_piezo}], whereas TLS1 gets symmetric at $39\,\mathrm{V}$ and TLS2 at around  $-10.8\,\mathrm{V}$. The top axes designate the corresponding value of $\varepsilon$. In Fig.~\eqref{fig_Kratio_meas}(b), we see an outlier value at $42\,\mathrm{V}$, which can be caused by a neighboring TLS that becomes resonant with the probed TLS at the chosen strain. The mean of the $K$-factor ratios is $2.5$ for TLS1 and $1.9$ for TLS2. Thus, we can estimate from Fig.~\eqref{fig_Kratio} that both TLS1 and TLS2 are positioned closer to the bottom electrode than to the top electrode (see both red horizontal lines labeled with TLS1 or TLS2). More precise elaboration of the QP penetration depth in the tunnel barrier [Appendix~\eqref{sec_QPTunnel}], as well as better estimation of the QP tunnel rate, would give a more concrete estimation of the TLS positions.\\
%--------------------SECTION----------------------------------------
\section{QP-induced pure dephasing rate of TLS}
\label{sec_TLSDephasing}
\begin{figure}[htbp]
	\begin{minipage}{0.49\textwidth}
%  	\psfragfig*[width=\linewidth]{./graphics/pureDephasing_vs_xqp/pureDephasing_vs_xqp}{
  	\includegraphics[width=\linewidth]{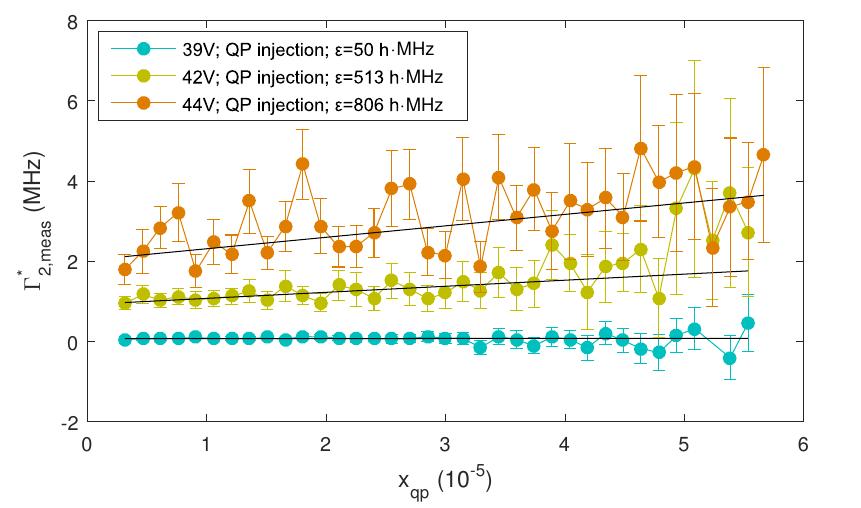}{
  	}
	\end{minipage}	
	\caption{Measured pure dephasing rate $\Gamma_{2,\text{meas}}^*$ of TLS2 vs density of injected quasiparticles $x_\text{qp}$ at various values of the asymmetry $\varepsilon$ (see legend). The data is fitted to the purely QP-induced dephasing rate shown in Eq.~\eqref{eq_gamma2TLS} (black lines). We clearly see that quasiparticle-induced pure dephasing of a TLS is enhanced when it is strain tuned away from the symmetry.}
	\label{fig_Gamma2}
\end{figure}

As shown in the previous section, the processes that allow QPs to be scattered into the opposite electrode contribute weakly to the TLS' decoherence. Hence, we use the simplified expression for QP-induced decoherence of the TLS shown in Eq. (5) in the main text to deduce the QP-induced pure dephasing rate $\Gamma_2^*$ by substituting the prefactor $\mathit{\Delta}/E$ by $\varepsilon/E$ and by setting $E\rightarrow 0$ in the integrand:
\begin{align}
\Gamma_2^*=&\;s_\text{B}^2\Gamma_{2}^{*\text{(B)}} + s_\text{T}^2\Gamma_{2}^{*\text{(T)}},\label{eq_gamma2TLS}\\
\Gamma_2^{*(l)}=&\;\frac{4\pi}{\hbar}\left(N_0Vg\frac{\varepsilon}{E}\right)^2\mathit{\Delta}_s\int_{1}^{\infty}d\epsilon\left(1-\frac{1}{\epsilon^2}\right)\notag\\
&\times \rho(\epsilon)^2 f_0^{(l)}(\epsilon)\left(1-f_0^{(l)}(\epsilon)\right).\label{eq_gamma2TLS_B}
\end{align}
In Fig.~\eqref{fig_Gamma2} we show the measured pure dephasing $\Gamma_{2,\text{meas}}^*$ of TLS2 in dependence of the injected QP density while it was strain tuned to various asymmetries $\varepsilon$ (see legend). The black lines are fits to the experimental data. As mentioned in the main text, the QP-induced dephasing increases with $x_\text{qp}$ when the TLS is strain tuned away from the symmetry, whereas it remains minimal for $\varepsilon\approx0$. Further, we recognize that the constant offset of the pure dephasing increases with $\varepsilon$ as it is dominated by interactions of the probed TLS with thermally fluctuating TLS~\cite{Lisenfeld2016}.\\

\clearpage
\bibliography{biblio}

\end{document}